\documentclass[aps,physrev,reprint,groupedaddress]{revtex4-2}

\usepackage{graphicx}
\usepackage{xcolor}
\usepackage{amsmath, amssymb}

\begin{document}


\title{Neural network expansion of Euclidean path integrals and its application to interacting scalar fields}


\author{G\'abor Balassa}
\affiliation{Department of Physics, Yonsei University, Seoul 03722, Korea}
\affiliation{Institute for Particle and Nuclear Physics, HUN-REN Wigner Research Centre for Physics, 29-33 Konkoly-Thege Mikl\'os \'ut, Budapest, 1121, Hungary}
\email[]{balassa.gabor@yonsei.ac.kr}

\date{\today}

\begin{abstract}
Studying phase transitions in interacting quantum field theories generally requires the numerical study of the dynamical system on a large lattice, which is, in most cases, computationally very challenging. In this work an alternative method is proposed to solve Euclidean path integrals in quantum field theories, using radial basis function-type neural networks. The method allows us to approximate observables in a very efficient manner, taking only seconds to do calculations that would otherwise take hours or even days with other existing methods. The model is used to describe phase transitions in the scalar $\phi^4$ theory for a wide range of coupling strength. The obtained phase transition line is compared to previous lattice results, giving very good agreement between them. 
\end{abstract}


\maketitle


\section{Introduction}
The solution of interacting quantum field theories generally poses great challenges even with modern numerical techniques and computational facilities. The problem lies in the fact that the interesting theories that correspond to real-world physics, like quantum chromodynamics \cite{1}, are nonlinear, containing complex self-interactions, and are generally very hard to grasp in an analytical fashion \cite{2}. In these cases one has to rely on numerical techniques, such as lattice methods \cite{3}, to solve the underlying field equations and to be able to give predictions for, e.g., particle masses \cite{4}, phase transitions \cite{5}, decay constants \cite{6}, potentials \cite{7}, etc. In many cases lattice methods are able to give very good results, however it generally takes a very long time (days, weeks, or even months) on supercomputers with direct GPU acceleration, parallelization, and specific hardware to achieve satisfying statistics \cite{8}. 

Another problem of lattice methods is the inability to efficiently simulate systems at finite densities due to the so-called sign problem \cite{9}. There are several attempts to overcome this by e.g. stochastic quantization and Langevin dynamics, reweighting techniques, imaginary chemical potentials etc. \cite{10,11,12}. While each of them has some success describing specific systems at specific densities, a fully satisfying description of finite density quantum chromodynamics is still missing. These issues make it necessary to work out alternate approaches that are capable of at least speeding up the calculations, but ultimately we would need an approach that is able to describe any nonlinear system efficiently even at finite densities. 

Recently, the application of various types of neural network configurations started to offer new ways to tackle problems in quantum mechanics, nuclear many-body physics, and field theories \cite{E1,E2,E3,E4,E5,E6,E7,E8,E9,E10,E11,15}.
One interesting approach could be the application of neural networks and machine learning techniques to approximate path integrals. In \cite{16} the Euclidean path integral formulation of quantum mechanics has been approximated by a multilayer perceptron-type neural network construction, where the nonlinear part of the action integral has been approximated by a sum of radial basis-type kernels, making the overall path integral analytically tractable. In this work, the method in \cite{16} is extended to be able to describe quantum field theories as well at arbitrary space-time dimensions, and is applied to the 1+1 dimensional interacting scalar field theory with a quartic self-interaction, that can be used to study spontaneous symmetry breaking, phase transitions, and critical phenomena \cite{17}. 

First, in Sec~\ref{sec:1} the Euclidean path integral formalism and the general method of the radial basis function expansion is described in detail. 
To show the working principles of the method, in Sec.~\ref{sec:2} it is applied to describe the coordinate space propagator of the free scalar field theory in 1+1 dimensions, while in Sec.~\ref{sec:22} the numerical complexity is briefly discussed.
After the free scalar field theory is discussed, in Sec.~\ref{sec:3} the RBF method is applied to the interacting $\phi^4$ theory, where the phase transition points for a wide range of bare couplings are derived and compared to previous lattice results. At the end we conclude the results and discuss further applicabilities and possible extensions of the model.

\section{General formalism}
\label{sec:1}
In quantum mechanics the path integral formalism provides a way to calculate quantum amplitudes by summing over all possible 'paths' a particle could take weighted by a factor that corresponds to the classical action integral of the underlying theory \cite{21}. In quantum field theory this idea is further generalized to fields, where instead of the $x(t)$ trajectories, one has to sum over all the possible $\phi(x)$ field configurations that satisfy the corresponding boundary conditions \cite{22}. The theory defined through path integrals is manifestly Lorentz invariant and has a deep connection to statistical mechanics through Wick rotating the time variable to imaginary times and turning the path integral into partition functions. In this work, only the Wick-rotated Euclidean formulation is considered, that in general can be used to calculate the masses of particles, critical points of phase transitions, fluctuations, or other thermodynamic observables. The general form of the Euclidean path integral formalism for real scalar fields can be written as:
\begin{equation}
\label{eq:1_1}
Z = \int \mathcal{D}\phi(x) \; e^{-\int \limits_{-\infty} \limits^{+\infty} d^Dx \; \mathcal{L}(\phi(x),\partial_{\mu}\phi(x))},
\end{equation}
where $\phi(x)$ represents the scalar field, $D$ is the dimension, while $\mathcal{L}$ is the Lagrangian density that describes the dynamics of the system.
The neural network method that will be used to approximate $Z$ is based on a radial basis function expansion of the nonquadratic (and possible quadratic) terms in the discretized action integral on a cubic lattice. 
In the following, we will set
$D=2$, and work out the method in 1+1 space-time dimensions, in which case we will have $ (x^0,x^1,x^2,...x^D) \rightarrow (t,x)$. 
This will not take away from the generality of the method, as the generalizations to higher dimensions are very straightforward and can be done easily by adjusting some parameters. The discretized version of the path integral in two dimensions can be written in a general form as:
\begin{equation}
\label{eq:26}
Z = \int  \prod_{i,j}d\phi_{i,j}  \; e^{ -a^2 \sum _{i,j}  L [\phi_{i+1,j},\phi_{i,j+1},\phi_{i,j} ] } ,
\end{equation}
where $\phi(t,x) \rightarrow \phi_{i,j}$, and the continuous field is approximated on an $N_t \times N_x$ lattice, with $\Delta t = \Delta x = a$ resolution. 
In general, assuming a forward numerical difference scheme in the kinetic terms, the Lagrangian will depend on the fields at the $(i,j)$ site and its neighbours as $L [\phi_{i+1,j},\phi_{i,j+1},\phi_{i,j} ]$.

The core idea of the radial basis function expansion is to stay with quadratic expressions at each lattice site. The first step of this is to separate the path integral into solvable and not solvable parts. In general this would mean the separation of quadratic and nonquadratic terms, however as we will see later, the restriction is not this severe and depends on the actual system under consideration. As a starting point, let us first separate purely the kinetic terms from the remaining parts that could include linear, quadratic, or higher-order terms.
In this work, we will only consider real scalar fields, however, the derivation to other types of fields follows the same method. 
By assuming periodic boundary conditions (i.e. $\sum_i \phi_{i+1} = \sum_i \phi_i$) the kinetic term can be given as:
\begin{eqnarray}
\label{eq:27}
\int d^2 x \; \left( \frac{1}{2} \partial_\mu \phi(x) \, \partial^\mu \phi(x) \right) \;  &\rightarrow& \; 
\frac{1}{2a^2} \phi^T M \phi,
\end{eqnarray}
where $a$ is the lattice resolution in both $x$ and $t$ directions, $N_t \times N_x$ is the number of lattice points, $M$ is an $N_tN_x \times N_t N_x$ matrix corresponding to the discrete Laplace operator, and $\phi=(\phi_{0,0},\phi_{0,1},...,\phi_{N_t-1,N_x-1})$ is an $N_t \times N_x$ vector. In this description $M$ does not depend on the step size $a$, which will be important later on. After separating the quadratic kinetic term from everything else, the discretized path integral can be written as follows:
\begin{equation}
\label{eq:28}
Z = \int \prod_{i,j}d\phi_{i,j}  \; e^{ -\frac{1}{2} \phi^T M \phi } \prod_{i,j}F [\phi_{i,j}] ,
\end{equation}
where $F [\phi_{i,j}]$ contains the possible nonlinear interaction as well as the quadratic mass terms in the case of massive fields and is given by the following form:
\begin{equation}
\label{eq:29}
F[\phi_{i,j}] = e^{  -a^2 L_I[\phi_{i,j}] } ,
\end{equation}
where $L_I[\phi_{i,j}]$ is again the remaining Lagrangian that is separated from the kinetic terms. 
In general this could include e.g. linear forcing terms $J_{i,j} \phi_{i,j}$, quadratic mass terms $m\phi_{i,j}^2$, or higher-order static nonlinearities and their functions like $\cos(\phi_{i,j}^2)$.
In the continuum theory this would correspond to a $F(\phi(x))$ term, that is a functional of the $\phi(x)$ field configurations. The main difference is that the corresponding discretized $F [\phi_{i,j}]$ is now a function of the $\phi_{i,j}$ field given at the $(i,j)$ space-time coordinate. Note that the $a$ dependence in the kinetic term vanished due to the double integral (double sum) in the action, which will not necessarily be true in other theories or in other dimensionalities. Finally, we have also used the fact that $\exp(\sum_i x_i) = \prod_i \exp(x_i)$ so that we will arrive at a factorized form at each lattice site, which will be important in the next steps.

Next, let us apply a radial basis function (RBF) expansion to the $F[\phi_{i,j}]$ function as:
\begin{eqnarray}
\label{eq:30}
F[\phi_{i,j}] \approx \sum_{k=1}^K  a_k e^{-b_k \Big(\phi_{i,j}-c_k \Big)^2 }= \\
\sum_{k=1}^K  a_k e^{ -A_k \phi_{i,j}^2 + B_k \phi_{i,j} - C_k }, \nonumber
\end{eqnarray}
where we have defined the new variables $A_k=b_k$, $B_k=2b_k c_k$, and $C_k=b_kc_k^2$ to separate the quadratic, linear, and constant shift terms. Using this expansion, the path integral can be written as a product of sums of $K$ quadratic kernels:
\begin{eqnarray}
\label{eq:31}
&&Z \approx \int \left( \prod_{i,j} d\phi_{i,j} \right) e^{  \frac{1}{2} \phi^T M \phi }    \times \\
&&\quad\quad\quad \prod_{i,j} \left[ \sum_{k=1}^K  a_k e^{  -A_k \phi_{i,j}^2 + B_k \phi_{i,j} - C_k }\right] , \nonumber
\end{eqnarray}
where, for now, we kept the kinetic terms (in its matrix form) separated from the RBF expansion. This form would have been very desirable if we hadn't mixed in the kinetic terms throughout the $M$ quadratic matrix, in which case the full integral could be factorized to ($K N_t N_x$) number of Gaussian integrals. Generally this would have been achievable by diagonalizing the $M$ matrix and going into momentum space, however, in this case this is not a straightforward task, mainly due to the linear shift in the RBF expansion that will bring in a mixing of the transformed field values. The next task is therefore to give an approximate solution to this problem, so that the path integral can be calculated in $\mathcal O(K \times N)$ complexity, where $N$ is the number of lattice points, and $K$ is the number of Gaussian kernels at each site.

First, let's rewrite the product of sums into a sum of $K^{N_t N_x}$ Gaussians as:
\begin{eqnarray}
\label{eq:32}
&& \prod_{i,j} \left[ \sum_{k=1}^K  a_k e^{   -A_k \phi_{i,j}^2 + B_k \phi_{i,j} - C_k}  \right]  = \nonumber \\
&=&   \sum_{\{k_{ij} \} \in K^{N_tN_x}}  \Bigg\{ 
\left( \prod_{i,j} a_{k_{i,j}} \right) \times \\ 
&&
e^{
    -  \sum_{i,j}  A_{k_{i,j}} \phi_{i,j}^2 
    +  \sum_{i,j}  B_{k_{i,j}} \phi_{i,j} 
    -  \sum_{i,j}  C_{k_{i,j}} 
} \Bigg\}, \nonumber
\end{eqnarray}
where we have a sum over all possible combinations between the expansions over all sites, and $k=\{k_{i,j}\}_{i=0,...,N_t-1,j=0,...,N_x-1}$, where each $k_{i,j} \in \{1,...,K\}$. The $A_{k_{i,j}}$, $B_{k_{i,j}}$, $C_{k_{i,j}}$, and $a_{k_{i,j}}$ parameters represent the original parameters in a specific $k_{i,j}$ combination. Putting back this expression into the path integral and writing everything in vector and matrix notation, we arrive at the following expression:
\begin{equation}
\label{eq:33}
Z \approx \!\!\!\!\!\!\sum_{\{k_{ij} \} \in K^{N_tN_x}} 
\!\!\!\!\!\! \hat{a}_{k}  \int \mathcal{D}\phi \; e^ { \frac{1}{2} \phi^T M \phi }
e^{
    - \phi^T \hat{A}_k \phi
    + \hat{B}_k^T \phi
    - 1^T \hat{C}_k
},
\end{equation}
where $ \mathcal{D}\phi=\prod_{i,j}d\phi_{i,j} $, $\hat{A}_k$ is an $N_t N_x \times N_t N_x$ diagonal matrix containing the $A_{k_{i,j}}$ elements, $\hat{B}_k$ and $\hat{C}_k$ are $N_t \times N_x$ vectors, $1^T$ is the corresponding unit vector, while $\hat{a}_{k}=\prod_{i,j} a_{k_{i,j}}$. In this case, we arrived at a sum of $K^{N_x N_t}$ general quadratic shifted path integrals that each have a closed-form solution, however, it quickly becomes unmanageable due to the large number of terms in the case of a larger lattice. In the next steps it will be shown that by a careful construction of the parameters it is possible to approximate this large sum by a much smaller number of terms in momentum space after diagonalizing the quadratic $M$ matrix.

\begin{figure}
    \includegraphics[width=.5\textwidth]{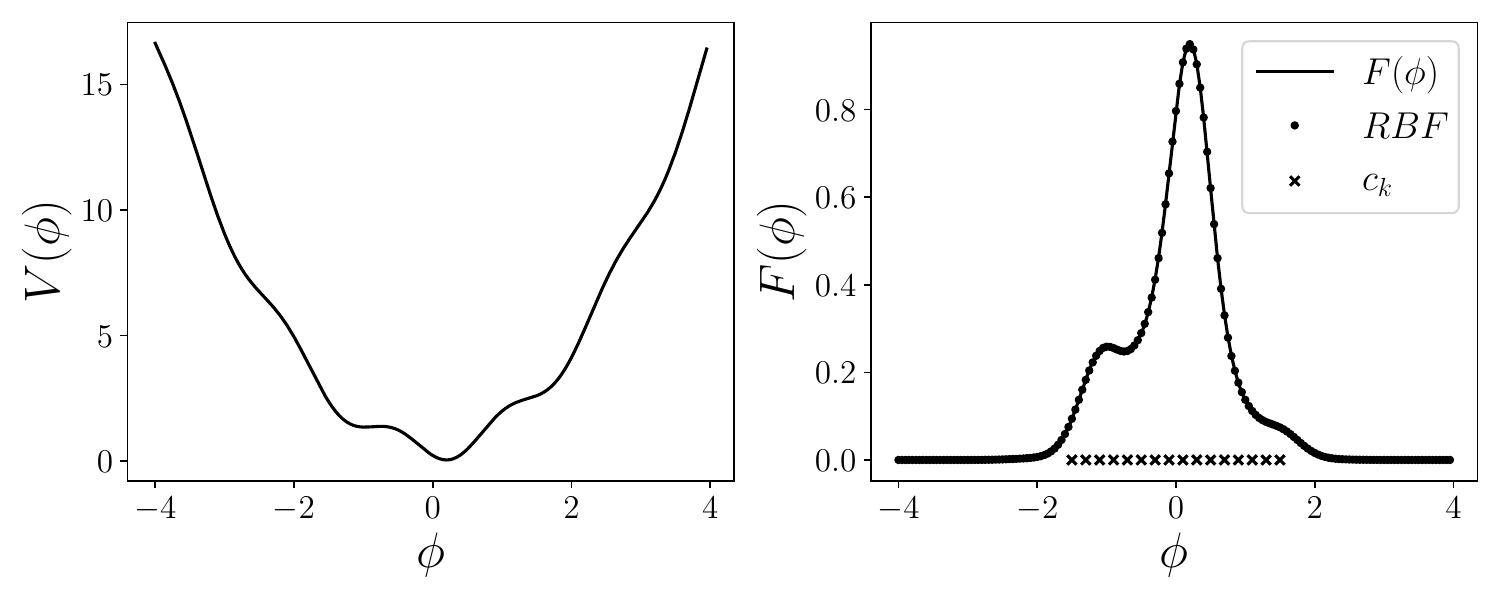}  
    \caption{RBF approximation of a test function $F(\phi)=e^{-V(\phi)}$, generated from the potential $V(\phi)=\phi^2 + \sin^2 \left( 2\phi-0.5 \right)$, with fixed Gaussian $A=5$ widths, and  K=16 uniformly distributed $c_k$ centers (shown with the 'x' symbols), between $-1.5$, and $1.5$ with $\Delta c_k=0.2$ resolution. }
    \label{fig:rbf2}
\end{figure}

The next step, therefore, is to simplify the form of the RBF expansion so that it is still able to approximate a large class of nonlinear functions. To do this, first, let us fix the $b_k$ width parameter of the Gaussians as $b_k = A$, which does not take away the generality of the RBF network and is, in many cases, a natural choice when one wants to achieve fast training. The centers $c_k$ and weights $a_k$ cannot be held constant, therefore, the corresponding $B_k$ and $C_k$ parameters in Eq.~\ref{eq:33} have to remain k-dependent. To show that this is indeed a valid choice, let us test it for a nonlinear $V(\phi)$ given as:
\begin{equation}
\label{eq:test}
V(\phi) = \phi^2 + \sin^2 \left( 2\phi-0.5 \right),
\end{equation}
where the corresponding $F(\phi)=e^{-V(\phi)}$ function will be the one that needs to be approximated by the RBF network.
Let us set the widths of the Gaussians to $A=3$, the centers to the interval $c_k \in [-1.5,1.5]$ with $\Delta c_k =0.2$, thus having $K=16$ number of kernels, and fit the $a_k$ parameters in a least square sense. The results can be seen in Fig.~\ref{fig:rbf2}, where the original function, the RBF approximation, and the centers are also shown.
It can be seen from this simple example that by setting the widths to a constant value and the centers to an interval that covers the range of the $F(\phi)$ function, it is possible to approximate any compact nonlinear function with very good accuracy. It is also worth noting that by an appropriate scaling of the function to a specific interval, the interval of the centers could also be held fixed, with the consequence that the scaling might change the necessary width of the Gaussians. Due to these reasons the determination of good $(A,c_k)$ parameters is not straightforward, but in practice does not pose any problems. 

In the next step, let us diagonalize the symmetric, circulant $M$ matrix that mixes the field components with its neighbors in coordinate space. To do this, we can apply a similarity transformation using a matrix $U$ to get $M=U\Lambda U^T$, where $\Lambda$ is now an $N_t N_x$ diagonal matrix, where the diagonal elements will be the eigenvalues of $M$. The $U$ transformation matrix is related to the discrete Fourier transform and can be built from the eigenvectors of $M$. By transforming the $\phi$ vectors with the unitary transformation $U$ as $\widetilde{\Phi} = U^T \phi$, the quadratic expression can be rewritten as $\phi^T M \phi \rightarrow \widetilde{\Phi}^T \Lambda \widetilde{\Phi}$, where $\Lambda=\text{diag}(\lambda_{ij})$. The values of $\lambda_{ij}$ can be given by the eigenvalues of the discretized Laplace operator that appeared due to the kinetic terms in the Lagrangian and can be written as:
\begin{equation}
\label{eq:34}
\lambda_{ij}= 4 \left[ \sin^2\left( \frac{p_i}{2}\right) + \sin^2\left( \frac{p_j}{2}\right) \right],
\end{equation}
where $p_{i(j)}=2 \pi n_{i(j)}/N_{t(x)}$ are the corresponding momenta, while $n_{i(j)} \in (-N_{t(x)}/2,N_{t(x)}/2]$, with $i(j) \in [0,N_{t(x)}-1]$.
Note that the eigenvalues of $M$ do not depend on the lattice resolution $a$, because it was separated from the matrix before. Now, let us apply the transformation to the other terms in Eq.~\ref{eq:32} that were coming from the RBF expansion, where the first purely quadratic term transforms as $\phi^T \hat{A}_k \phi \rightarrow A \widetilde{\Phi}^T\widetilde{\Phi}$, due to the fact that we have assumed a constant $A$ width parameter for the Gaussian kernels. The constant shift $\hat{C}_k$ does not transform because it is not coupled to the fields, however, the linear shift will become nontrivial because $\hat{B}_k^T \phi \rightarrow \hat{B}_k^T U \widetilde{\Phi} = (U^T \hat{B}_k)^T \widetilde{\Phi}$. Therefore, the transformation will mix the $\widetilde{\Phi}$ terms through the $U$ transformation matrix, or, by looking at it another way, transforms the $B_k$ parameter vectors. Writing back everything in Eq.~\ref{eq:32} the path integral becomes:
\begin{equation}
\label{eq:35}
Z \approx \!\!\!\!\!\! \sum_{\{k_{ij} \} \in K^{N_tN_x}} 
\!\!\!\!\!\! \hat{a}_{k}  \int \mathcal{D}\widetilde{\Phi} \; e^{  \frac{1}{2} \widetilde{\Phi}^T \Big[\lambda-2A\Big] \widetilde{\Phi}
    + (U^T\hat{B}_k)^T \widetilde{\Phi}
    - 1^T \hat{C}_k
},
\end{equation}
where $\widetilde{\Phi}$ is the momentum space field vector. This is now in a diagonal form and can be integrated out easily to get:
\begin{equation}
\label{eq:36}
Z  \approx  \!\!\! \!\!\! \sum_{\{k_{ij} \} \in K^{N_tN_x}} \!\!\!\!\!\! \hat{a}_{k} \,\, \prod_{i,j }\left( \frac{2 \pi}{\lambda_{ij}+2A} \right)^{\frac{1}{2}} e^ { \frac{(U^T \hat{B}_k)_{i,j}^2}{2 (\lambda_{ij}+2A) } - C_{k_{i,j}}},
\end{equation}
where $(U^T \hat{B}_k)_{i,j}$ is the $(i,j)$'th element (e.g. in a lexicographic ordering) of the transformed $\hat{B}_k$ vector. By substituting back the original parameters: linear weights $a_k$, centers $c_k$, and widths $b_k=A \; (\forall k)$ of the RBF network, the path integral can be written as:
\begin{equation}
\label{eq:37}
Z \approx \!\!\! \!\!\! \sum_{\{k_{ij} \} \in K^{N_tN_x}}   \prod_{i,j} a_{k_{i,j}} \left( \frac{2 \pi}{ \lambda_{ij}+2A} \right)^{\frac{1}{2}} e^{ \frac{2A^2(U^T \hat{c}_k)_{i,j}^2}{(\lambda_{ij}+2A) } - A c_{k_{i,j}}^2} ,
\end{equation}
where $\hat{c}_k$ is a vector of the RBF centers corresponding to the $k$'th combination, while $c_{k_{i,j}}$ is one element of this vector. 

The path integral in this representation after diagonalization still contains a very large sum of weighted Gaussian path integrals, but now the problem is shifted from the mixing of the field values through the kinetic terms in coordinate space to the mixing of the predefined centers of the kernel functions in the linear terms in momentum space. This in itself still makes it impossible to rearrange the path integral into a product of factorized quadratic path integrals. It turns out, however, that by a careful construction of the RBF parameters, the full sum can be approximated by a fully separable model, where the mixing disappears and the full path integral could be written as:
\begin{equation}
\label{eq:S}
Z \rightarrow \prod_{i,j} \sum_{k=1}^K Q_k \Big[ \widetilde{\Phi}_{i,j},\lambda_{ij},A, c_k,a_k \Big].
\end{equation}

The problematic part that is in the way of achieving this form is the transformed $(U^T \hat{c}_k)$ centers of the Gaussian kernel functions. If we could write this as simply $\hat{c}_k$, without the transformation, then the whole expression in Eq.~\ref{eq:37} could be written in the form of Eq.~\ref{eq:S}. In general this is not possible, however, we see that in Eq.~\ref{eq:37} the path integral depends on a nontrivial expression of the $A$ parameters, the $\lambda_{ij}$ eigenvalues, and the $c_k$ centers, therefore, it could be possible to select such parameters that make the difference between the path integrals, with and without transforming the centers, negligible, while still making it possible to estimate any nonlinear relationship. To quantify this, let us set $a_k=1 \; (\forall k)$ and define $Z_0$ and $Z_1$ as follows:
\begin{equation}
\label{eq:38}
	Z_0(N_C) = \!\!\!\! \sum_{\{k_{ij} \} \in N_C}   \prod_{i,j} \left( \frac{2 \pi}{ \lambda_{ij}+2A} \right)^{1/2} e^ { \frac{2A^2(U^T \hat{c}_k)_{i,j}^2}{ (\lambda_{ij}+2A) } - A c_{k_{i,j}}^2 } ,
\end{equation}
\begin{equation}
\label{eq:39}
Z_1(N_C)= \!\!\!\! \sum_{\{k_{ij} \} \in N_C} \prod_{i,j} \left( \frac{2 \pi}{ \lambda_{ij}+2A} \right)^{1/2} e^{ \frac{2A^2c_{k_{i,j}}^2}{ (\lambda_{ij}+2A) } - A c_{k_{i,j}}^2}  ,
\end{equation}
where $Z_0$ is the full expression including the transformed centers, while $Z_1$ is the path integral where the centers are not transformed. The $N_C$ parameter is the number of terms that we will consider in the summation, which generally should be smaller than all the possible combinations when we choose large $N_t$ and $N_x$, therefore, to make the calculations numerically tractable, we will consider $N_C < K^{N_t N_x}$. Statistically this will not be a problem if we choose large enough $N_C$ and use sufficiently enough samples $M$ to make the simulations. This will be clarified in more detail when we define the algorithm to obtain the error distributions.

The task is to compare $Z_0$ and $Z_1$ for different $c_k$ and $A$ parameters and determine the cases where the differences are negligible. Considering the fact that $Z$ in itself is not a meaningful quantity and could take very large or very small values, the necessary quantity that is used to calculate the observables and that we would need to compare is actually $W=\ln Z$, therefore, the error measure we will use is defined as the averaged absolute relative error between the logarithm of $Z_0$ and $Z_1$:
\begin{equation}
\label{eq:40}
E_R(\ln Z) = \frac{1}{M} \sum_{i=1}^M \left[ \frac{|\ln Z_{0,i}(N_C) - \ln Z_{1,i}(N_C)|}{|\ln Z_{0,i}(N_C)|} \right],
\end{equation}
where $M$ is the number of samples, each having $N_C$ different randomly generated combinations of the kernels that are defined for fixed centers and widths.
This error definition is meaningful because right now, we are not interested in pointwise accuracy, but rather in the resulting distribution of the original and transformed points, that are part of the large sum. At the end the mean of these distributions will correspond to the full sum, or in other words to the result of the path integral. 

To estimate the error in $\ln Z$ that is induced by the unitary transformation, the following steps have been followed:
\begin{itemize}
\item Fix the parameters of the RBF network, which include the number of kernels $K$, the width of the Gaussians $A$, and the k-number of centers $c_k$. The centers are chosen in a predefined interval $[a,b]$, with some $\Delta c_k$ resolution.
\item Set the lattice size $N_t$ and $N_x$, the number of combinations $N_C$, and the number of samples $M$.
\item Generate $N_C$ randomly chosen combinations from all the possibilities. By doing this $M$ times, we can obtain the statistics of the error distributions, where $E_R$ is defined as the mean value of the obtained errors in the different samples.
\end{itemize}

\begin{figure}
    \centering
    \includegraphics[width=0.47\textwidth]{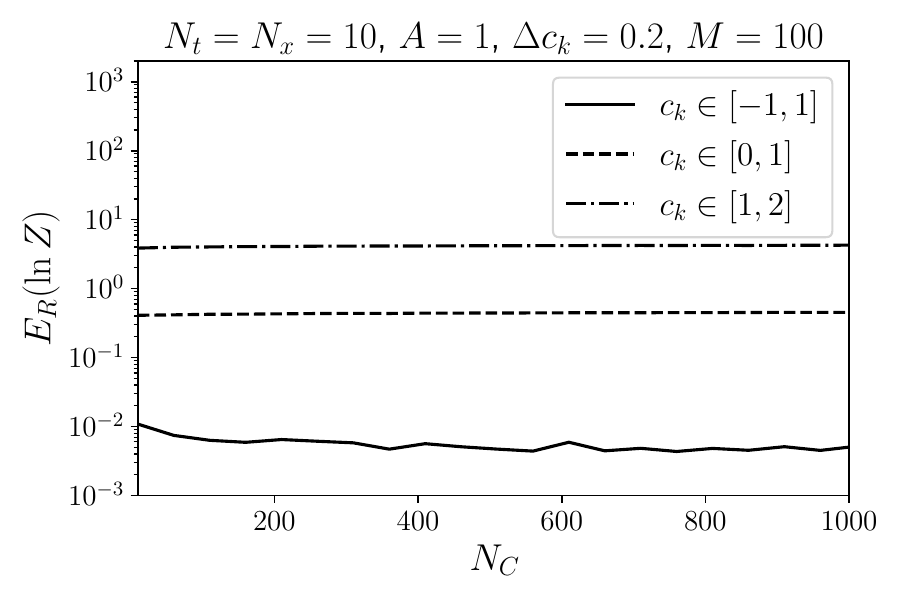}  
    \caption{Dependence of the relative error, defined in Eq.~\ref{eq:40}, on the number of combinations $N_C$ for different sets of $c_k$ centers for fixed $A=1$ widths. The corresponding parameters of the system and the RBF network are given in Tab.~\ref{tab:1}.}
    \label{fig:S1}
\end{figure}

For the first test let us compare the $N_C$ dependence of the $E_R(\ln Z)$ error in three different scenarios where the parameters are given in Tab.~\ref{tab:1}. The aim in this simulation is to see how the error evolves with the number of $N_C$ combinations and to give a first estimate of the magnitudes of the error when the centers are distributed symmetrically around 0 and in the case when they are not symmetric to 0. The result can be seen in Fig.~\ref{fig:S1}, where some very clear conclusions can be made by examining the errors.

\begin{table}[h!]
\centering
\begin{tabular}{p{1.5cm}|p{1.5cm}|p{1.5cm}|p{1.5cm}|p{1.5cm}}
 \hline
 $M$ & $A$ & $c_k$ & $\Delta c_k$ & $N_t, N_x$ \\
 \hline\hline
 $100$ & $1$ & $[-1,1]$ & $0.2$ & $10,10$ \\
 $100$ & $1$ & $[0,1]$ & $0.2$ & $10,10$ \\
 $100$ & $1$ & $[1,2]$ & $0.2$ & $10,10$ \\
\end{tabular}
\caption{Parameters for the first test to observe the relative error dependence on the number of combinations $N_C$ for three different RBF networks. The corresponding results are shown in Fig.~\ref{fig:S1}.}
\label{tab:1}
\end{table}

The most important observation we could make is the difference in the magnitude of the errors for the symmetric and for the nonsymmetric cases, where, in the case of symmetric centers, the error is much smaller than in the two other cases. The other observation is that the relative error tends to be saturating after taking a few thousand combinations, which property makes the use of $N_C < K^{N_t N_x}$ a sensible choice. The number of combinations that it takes to achieve the saturation depends on the distribution of the specific centers.

\begin{figure}
    \centering
    \includegraphics[width=0.47\textwidth]{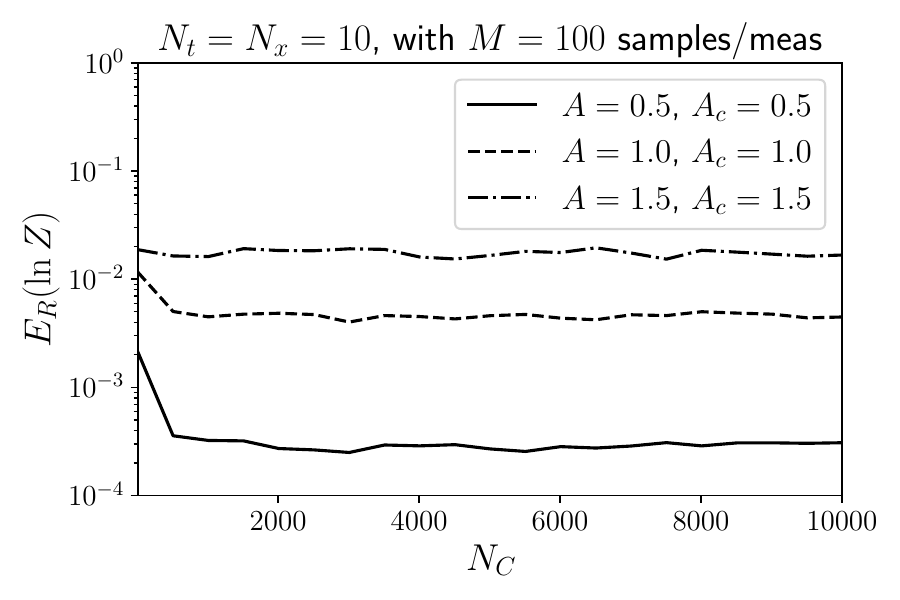}  
    \caption{The $N_C$ (number of combination) dependence of the relative errors, defined in Eq.~\ref{eq:40}, for different sets of RBF parameters using only symmetric centers around zero, where $A_c$ represents the boundary values of the centers as $c_k \in [-A_c,A_c]$. The corresponding parameters are given in Tab.~\ref{tab:2}.}
    \label{fig:S2}
\end{figure}

In the second example let us further analyze the $N_C$ dependence of the relative errors but now only using symmetric centers, with different width parameters $A$, and with different intervals defined as $c_k \in [-A_c, A_c]$, with $K=10$ and $\Delta c_k = 0.2 A_c$. The parameters are collected in Tab.~\ref{tab:2}, while the results are shown in Fig.~\ref{fig:S2}.

\begin{table}[h!]
\centering
\begin{tabular}{p{1.5cm}|p{1.5cm}|p{1.8cm}|p{1.5cm}|p{1.5cm}}
 \hline
 $M$ & $A$ & $c_k$ & $\Delta c_k$ & $N_t, N_x$ \\
 \hline\hline
 $100$ & $0.5$ & $[-0.5,0.5]$ & $0.1$ & $10,10$ \\
 $100$ & $1$ & $[-1,1]$ & $0.2$ & $10,10$ \\
 $100$ & $1.5$ & $[-1.5,1.5]$ & $0.3  $ & $10,10$ \\
\end{tabular}
\caption{Parameters for the second test to observe the relative error dependence on the number of combinations $N_C$ for three different RBF networks using only symmetric $c_k$ centers around zero. The corresponding results are shown in Fig.~\ref{fig:S2}.}
\label{tab:2}
\end{table}

The results shown in Fig.~\ref{fig:S2} show the same behavior as what we have observed previously in Fig.~\ref{fig:S1}, where the symmetric center configuration was able to give a very small relative error in the range of a few percentage. The $N_C$ dependence also shows the same behavior, where after a few hundred combinations the relative error tends to saturate, therefore, we do not need to consider all the possible combinations to be able to say something about the error distribution. The important takeaway from the results in Fig.~\ref{fig:S2} is that the magnitude of the error depends on the distribution of the centers and also on the width of the Gaussian kernels, which is expected as $A$ is also present in $Z_0$ and $Z_1$.

\begin{figure}
    \centering
    \includegraphics[width=0.5\textwidth]{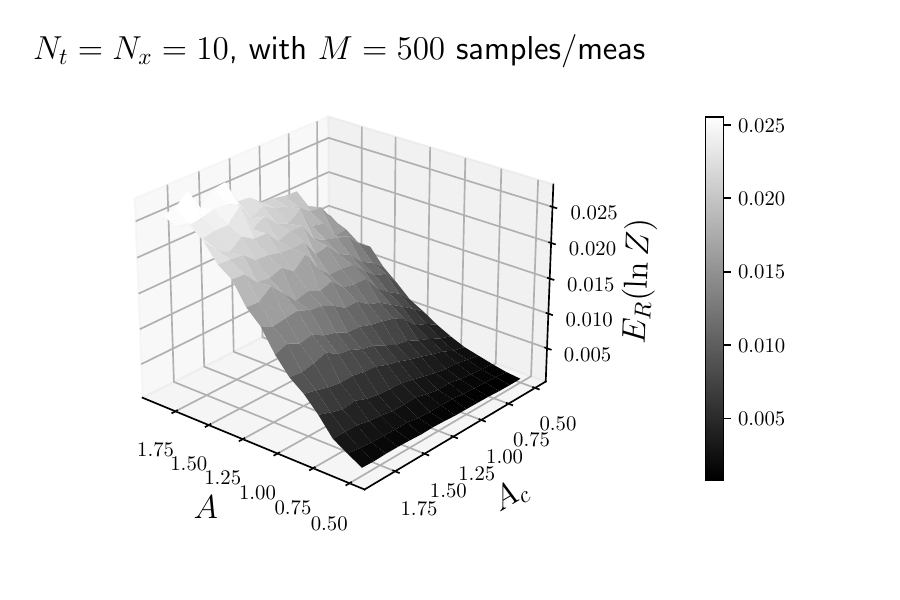}  
    \caption{Relative error dependence on the Gaussian width $(A)$ and on the symmetric center boundary $(A_c)$ parameters, where the centers are defined as $c_k \in [-A_c,A_c]$. To reach saturation, $N_C=200$ number of combinations were used in each sample, and to get good statistics, $M=500$ samples were used per measurement. }
    \label{fig:S3}
\end{figure}

In Fig.~\ref{fig:S3} we further analyze the dependence of the relative error on the $A$ and $c_k$ parameters, where again only symmetric centers were used with a sensible range given by $[-A_c,A_c]$ with $A_c \in [0.5,2]$ and $A \in [0.5,2]$. In these calculations the number of combinations is fixed to $N_C=200$, so that a saturation of the relative errors is reached, and we take $M=500$ samples per measurement to get good statistics.

From the results shown on Fig.~\ref{fig:S3}, it can be seen that the relative error stays under a few percent in the intervals that were given. This is a general behavior of the error, and the most important result we can deduce from these simulations is that the relative error in $\ln Z$ stays very low in the case when we use $c_k$ centers that are in a well-defined range that is symmetric around zero. The error depends on the width parameters as well, however, for sensible choices the error still stays under a few percent. The actual parameters that are needed to approximate a specific function depend heavily on the function itself, but this function can always be scaled and shifted so that it is centered around zero. This does not change the observables that can be extracted from the path integrals, therefore, selecting a symmetric $c_k$ is always possible. Special care might be needed in the case of gauge theories with local symmetries, however, in the case of the interacting scalar fields, it is not a problem. The value of $A$ depends on the shape of the function, e.g., if it's heavily oscillating, a larger $A$ could be needed. For smooth, nonoscillating functions, choosing $A \in \mathcal{O}(1)$ is a good starting point.

Next, let us also examine the actual distribution of the errors with a fixed $K=20$ number of kernels, $c_k \in [-A_c,A_c]$ centers, and different $A$ and $A_c$ intervals shown in Tab.~\ref{tab:3}. The simulations are done by specifying one $A$ width and one $A_c \rightarrow c_k \in [-A_c,A_c]$ center (from the previously determined intervals for $A$ and $A_c$) at each sample, calculating the errors for these values, and then doing this $M$ times. After we have enough statistics, the histogram of the errors can be obtained. This is also an important quantity, as it is needed when we want to, e.g., calculate the uncertainty of the observables that are calculated with the RBF model. In these simulations the relative errors and the normalized errors defined in Eq.~\ref{eq:E2} and Eq.~\ref{eq:E3} are both calculated using $M=10000$ samples to estimate the histograms.

\begin{equation}
\label{eq:E2}
E_{rel} = \frac{Z_0-Z_1}{Z_0}, 
\end{equation}
\begin{equation}
\label{eq:E3}
E_{normalized} = \frac{Z_0-Z_1}{\max{ \{ |Z_{0,i}-Z_{1,i}|_{\forall i}}\} },
\end{equation}
where $\max{ \{ |Z_{0,i}-Z_{1,i}|_{\forall i}}\}$ is the maximum value of the absolute differences, chosen from all of the samples. The normalization in $E_{normalized}$ is used because only the shape of the distribution is important and not the actual value of the difference.

\begin{figure}
    \centering
    \includegraphics[width=0.48\textwidth]{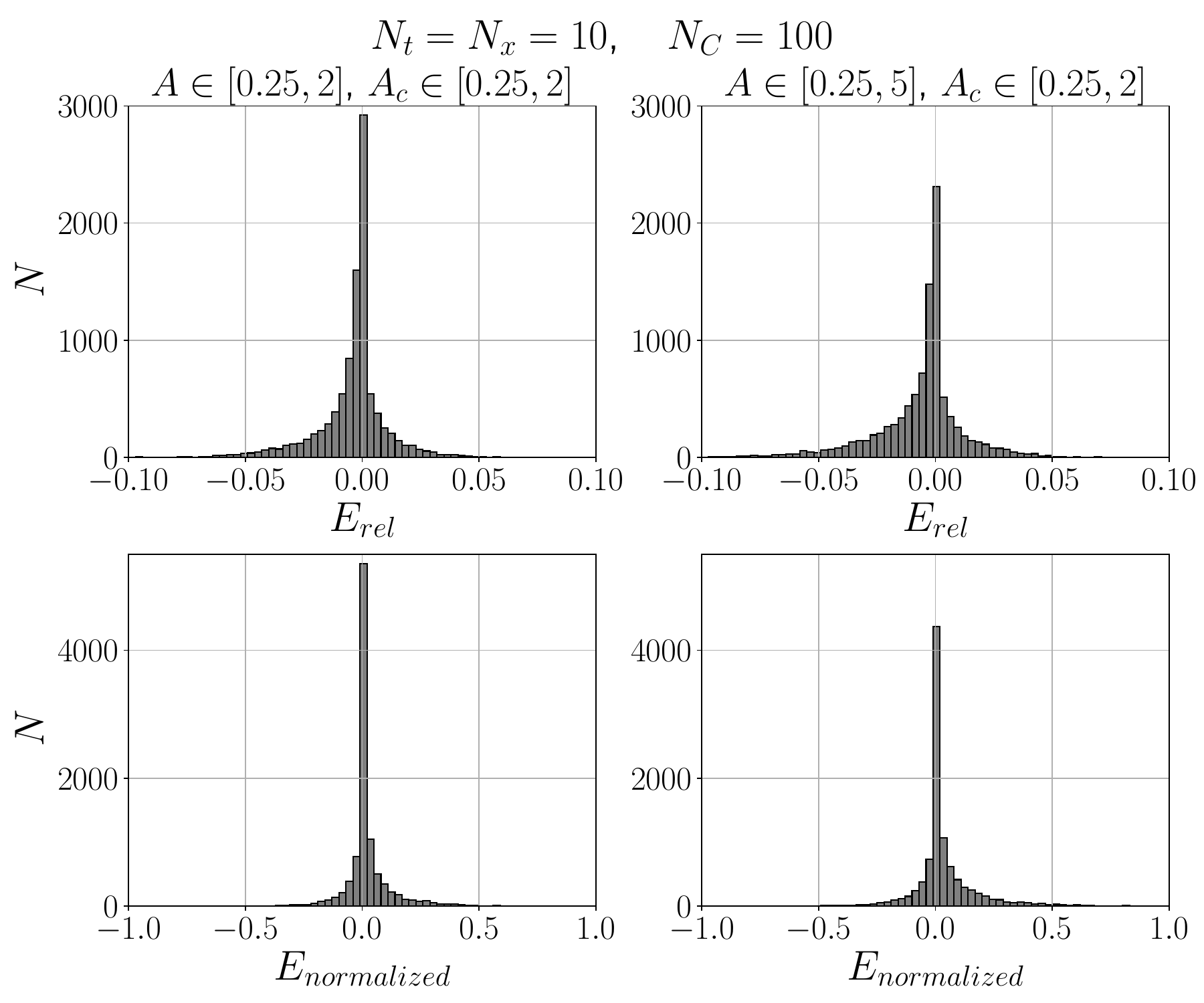}  
    \caption{Distributions of the errors defined in Eq.~\ref{eq:E2} and Eq.~\ref{eq:E3} using the parameters summarized in Tab.~\ref{tab:3}. On the upper two plots the distribution of the relative errors is shown, while on the bottom two plots the distribution of the normalized difference is shown.}
    \label{fig:S4}
\end{figure}

\begin{table}[h!]
\centering
\begin{tabular}{p{1.5cm}|p{1.5cm}|p{1.5cm}|p{1.5cm}}
 \hline
 $A$ & $A_c$ & $K$ & $N_t, N_x$ \\
 \hline\hline
 $[0.25,2]$ & $[0.25,2]$ & $20$ & $10,10$ \\
 $[0.25,5]$ & $[0.25,2]$ & $20$ & $10,10$ \\
\end{tabular}
\caption{Parameters used to examine the distributions of the errors defined in Eq.~\ref{eq:E2} and Eq.~\ref{eq:E3} for a fixed $A_c$ interval and two different $A$ interval sets. The parameter intervals mean that a specific parameter combination is selected from the given intervals in each sample. The corresponding results are shown in Fig.~\ref{fig:S4}.}
\label{tab:3}
\end{table}
In Fig.~\ref{fig:S4} the distribution of the relative error and the normalized difference between $Z_0$ and $Z_1$ can be seen, showing a clear peak near zero, which is the expected behavior for symmetric centers.

The distributions also show that for smaller $A$ parameters, the peak near zero is larger, which means that the approximation will be better by using smaller widths for the same symmetric centers.
After observing the behavior of the relative errors under specific parameter sets, we could deduce, that if we set the $c_k$ centers to be symmetric around zero, with some corresponding $A$ width parameter, then the path integral can be approximated by a separable form in momentum space that is shown in Eq.~\ref{eq:S} with an accuracy of a few percent for $\ln Z$.

The achieved accuracy shown in these examples is a property of the approximation strictly between $Z_0$ and $Z_1$ and does not represent the full accuracy of the RBF approximation of the path integral, which also depends on the accuracy between the applied RBF network and the $F(\phi)$ function. In general, the former gives an upper limit that depends on the number and distribution of the centers and the width parameters and cannot be improved systematically by, e.g., introducing more kernels. On the other hand, the latter can be improved by, e.g., using more kernels at specific parts, where the $F(\phi)$ function might need a better resolution. Due to the universal approximation property of the RBF networks, in theory this could be improved 'indefinitely'. The full accuracy of the RBF approximation, however, will still depend on both terms, therefore, it cannot be systematically improved to achieve a perfect approximation. 

The simulations shown here have been done in 1+1 dimensions, however, the situation does not change in higher dimensions. The actual values of the centers and widths and the corresponding relative errors will change, but the original statement, as the path integral could be approximated by a good accuracy, stays the same.

\begin{figure}
    \centering
    \includegraphics[width=0.48\textwidth]{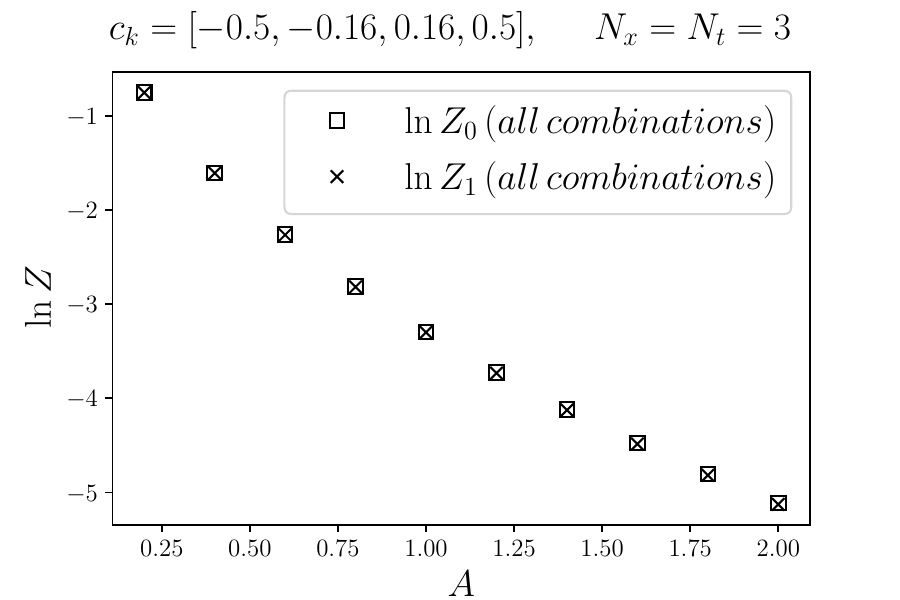}  
    \caption{Comparison of the full RBF sum ($Z_0$) and its approximation ($Z_1$) on a $3 \times 3$ lattice, with $K=4$ number of centers using different Gaussian width parameters. The full $K^{N_tN_x}$ sum shows the same behavior and accuracy as it was shown before with $N_C < K^{N_tN_x}$ number of samples.}
    \label{fig:FULL}
\end{figure}

Lastly, let us also show on a tractable example that the approximation is valid even when we include all the possible combinations in the $K^{N_tN_x}$ sum. Let us set up $K=4$ centers in a symmetric fashion around zero as $c_k=[-0.5,-0.16,0.16,0.5]$ and compare $\ln Z_0$ and $\ln Z_1$ using all the combinations on a $3 \times 3$ lattice, in which case the number of terms will be $4^{9}=262144$. The results can be followed in Fig.~\ref{fig:FULL}, where the calculations have been done for a set of width parameters ($A$) between $0.2$ and $2$. The results show a very good (under a percent) accuracy, which is consistent with our previous calculations using $N_C<K^{N_tN_x}$ samples.

To summarize the method in 1+1 dimensions, let us write down the steps and the resulting approximation for the path integral in Eq.~\ref{eq:41}:
\begin{eqnarray}
\label{eq:41}
&&Z \approx \sum_{\{k_{ij} \} \in K^{N_tN_x}} 
\hat{a}_{k}  \int \mathcal{D}\widetilde{\Phi} \; e^{ \frac{1}{2} \widetilde{\Phi}^T \Big[\Lambda-2A\Big] \widetilde{\Phi}
    + (U^T\hat{B}_k) \widetilde{\Phi}
    - 1^T \hat{C}_k
}  \nonumber \\
&&
\rightarrow \text{ (choose good $c_k$ and $A=b_k (\forall k)$,  RBF parameters)}   \nonumber \\
&&\rightarrow
\sum_{\{k_{ij} \} \in K^{N_tN_x}} 
\hat{a}_{k}  \int \mathcal{D}\widetilde{\Phi} \; e^{ \frac{1}{2} \widetilde{\Phi}^T \Big[\Lambda-2A\Big] \widetilde{\Phi}
    + \hat{B}_k \widetilde{\Phi}
    - 1^T \hat{C}_k 
}    \\
&&\rightarrow 
\prod_{i,j} \int d\widetilde{\Phi}_{i,j} \;  \left[ \sum_{k=1}^K  a_k e^{   -\frac{1}{2} \Big[ \lambda_{ij}-2A\Big] \widetilde{\Phi}_{i,j}^2 + 2Ac_k \widetilde{\Phi}_{i,j} - A c_k^2 } \right] 
, \nonumber
\end{eqnarray}
where the first line shows the large sum of quadratic path integrals in momentum space, while the third line shows its approximation with a good RBF parametrization. In the fourth line, we have rewritten the path integral into a product of sums that is factorized at the different lattice sites. Using this representation, the integral can be carried out and has the closed-form as follows:
\begin{equation}
\label{eq:42}
Z \approx \prod_{i,j} \left[ \sum_{k=1}^K a_k \left( \frac{2 \pi}{ \lambda_{ij}+2A} \right)^{\frac{1}{2}} e^{  \frac{2 A^2 c_k^2}{ (\lambda_{ij}+2A) } - A c_k^2 } \right] .
\end{equation}

Using this approximation, the original $\mathcal{O}(K^{N_t  N_x})$ complexity is reduced to $\mathcal{O}(K  N_t  N_x)$ when we want to calculate the full path integral. 

Before we move on to concrete examples, two remarks are in order concerning the selection of the centers and the scaling of the fields. Generally the $F[\phi]$ functions that we want to estimate by the RBF network tend to go to zero much faster than the corresponding radial basis functions, therefore, one has to be careful with centers that are far from the dominant parts of the original function, as they could include extra contributions to the path integral in regions that otherwise could have a larger suppression. In practice, it is enough to make sure that the centers are positioned in an interval where the $F[\phi]$ function has dominant contributions. To make sure that we did not include extra contributions, it is advisable to do the calculations with many different parametrizations and compare the results for the observables. Another possible method could be the use of definite integrals of the RBF approximation in a bounded region, where the magnitude of $F[\phi]$ is non-negligible, which could also be expressed in a closed form due to the Gaussian integrals that are involved in the path integral. In this work we will use the former method and express the integrals on an infinite range.

The last remark is about the scaling properties of the fields. By using a simple $\phi \rightarrow S_c\phi$ scaling transformation, the function that needs to be approximated on every lattice site also changes as $F[\phi] \rightarrow F[S_c \phi]$. In this way $F[\phi]$ can be shaped so that a good RBF parametrization can be used, therefore, the factorized form in Eq.~\ref{eq:S} can be achieved. The scaling also changes the kinetic terms as $\frac{1}{2}\phi^T M \phi \rightarrow \frac{1}{2}\phi^T (S_c^2 M)\phi$, so the $M$ matrix gets an extra $S_c^2$ factor, in which case the eigenvalues also change to $\lambda \rightarrow S_c^2 \lambda$. Lastly, the integral measure also gets an extra factor as $d\phi \rightarrow S_c d\phi$, so the final result for the integrated path integral can be written as:
\begin{equation}
\label{eq:43}
Z(S_c) \approx S_c \prod_{i,j} \left[ \sum_{k=1}^K a_k \!\left( \frac{2 \pi}{ S_c^2 \lambda_{ij}+2A} \right)^{\frac{1}{2}} \!\! e^{  \frac{2 A^2 c_k^2}{(S_c^2 \lambda_{ij}+2A) } - A c_k^2 } \right] ,
\end{equation}
where we have accounted for all the extra $S_c$ scaling factors. In the next section the method is applied to the system of free scalar fields, to show its working principles on an analytically tractable example.

\section{Example: the free scalar field in 1+1 dimensions}
\label{sec:2}
The noninteracting real scalar field theory in 1+1 dimensions is the simplest system where the method could be easily followed. The calculation for more complex theories with nonlinear interactions at higher dimensions follows the same procedure as it will be shown here. On a side note, in the case of gauge theories, or theories including Grassman variables, e.g., for fermionic fields, the method needs to be extended due to the different kinetic terms, different symmetries, noncommutativity of field variables etc. 
In the case of geometrically non-flat field spaces, e.g., in gauge theories or sigma models, simple Gaussian kernels might not be enough, and other types of kernels and/or specific transformation rules applied to the RBF networks could be needed. We will leave these complications for future work.

In this section, we will show the working principles of the method by using three different parametrizations of the RBF networks, while also including one instance where the field is scaled to a smaller interval using $S_c > 1$. For now, staying with the problem at hand, the Lagrangian density for the free scalar field with mass $m$ can be written as:
\begin{equation}
\label{eq:F1}
\mathcal{L}= \frac{S_c^2}{2} \partial_\mu \phi(x) \, \partial^\mu \phi(x) + S_c^2\frac{m^2}{2} \phi(x)^2.
\end{equation}
After discretizing the corresponding action in $1+1$ dimension with $\Delta t = \Delta x = a$, and separating the kinetic term from the quadratic parts, we define the $F(\phi_{i,j})$ function (that has to be approximated by the RBF network) as follows:
\begin{equation}
\label{eq:F3}
F(\phi_{i,j}) = e^{ - S_c^2\frac{(am)^2}{2}\phi_{i,j}^2  },
\end{equation}
where we have kept the dependence on the $S_c$ scaling and $a$. In order to show that different good parametrizations will give the same results, we have set $a=1$, and done the calculations using three different RBF networks with the parameters shown in Tab.~\ref{tab:4}.

In all cases we set the lattice size to $N_x=N_t=40$, and the mass to $m=1$, in which case the $F(\phi)$ function will be dominant in the range of $\phi \in [-4,4]$. In the first parametrization we have applied a scaling of $S_c=4$, which scales the function into the new range of $\phi^* \in [-1,1]$. This makes it possible to use smaller centers near zero in the account of a possibly larger $A$ Gaussian width parameter. In the second case no scaling was applied for the fields with uniformly distributed centers around zero. In this case a smaller $A$ parameter could be used due to the larger interval that $F[\phi]$ covers. In the third case, similar to the second one, no scaling was applied, however, here the centers have been chosen randomly and not distributed uniformly around zero. 
Note that in all cases the $c_k=0$ center is omitted, so that the training does not try to fit the RBF network dominantly with the $c_k=0$ kernel. In the case of $A=\frac{S_c^2a^2m^2}{2}$ the network would become trivial with only the $c_k=0$ having $a_k=1$ nonzero weight.
By omitting this term, we force the network to try to estimate the nonlinear function with a set of shifted Gaussians, with different, nonzero weights in each case. 

\begin{figure}[htbp]
    \centering
    \includegraphics[width=0.5\textwidth]{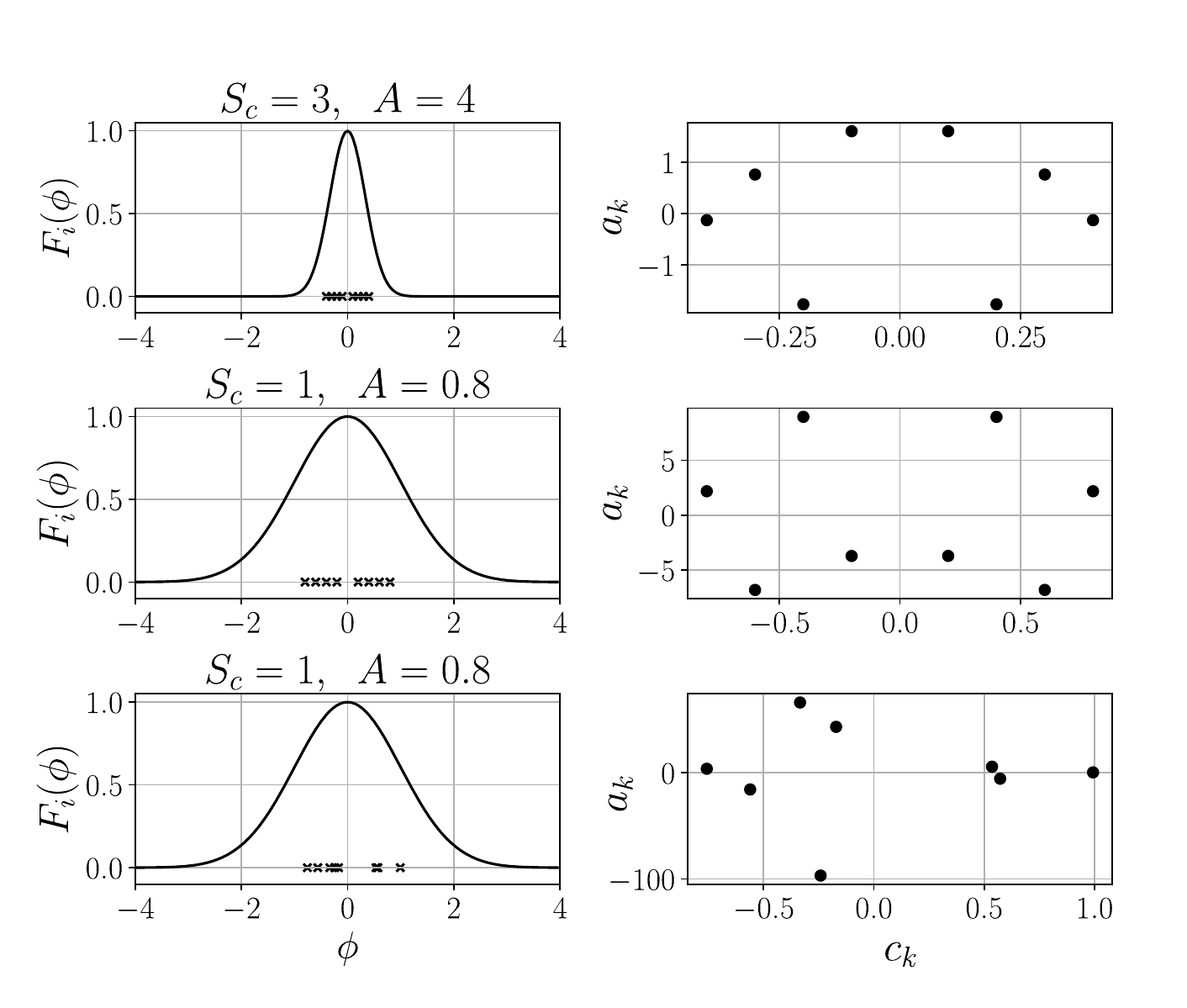}  
    \caption{RBF approximation of the $F(\phi)$ function that is defined in Eq.~\ref{eq:F3} for the free scalar field case. The left-hand side plots show the RBF approximations of the (scaled) $F(\phi)$ functions with the positions of the RBF centers, while the right-hand side plots show the fitted $a_k$ RBF weights for each case.}
    \label{fig:SC1}
\end{figure}

The results can be seen in Fig.~\ref{fig:SC1}, where the RBF approximation of the $F(\phi)$ functions and the trained $a_k$ parameters are also shown for the three different parametrizations.

\begin{table}[h!]
\centering
\begin{tabular}{p{0.4cm}|p{7.2cm}|p{0.5cm}}
 \hline
 $S_c$ & $c_k$ & $A$  \\
 \hline\hline
 $3$ & $[-0.4,-0.3,-0.2,-0.1,0.1,0.2,0.3,0.4]$ & $4$ \\
 $1$ & $[-0.8,-0.6,-0.4,-0.3,0.3,0.4,0.6,0.8]$ & $0.8$  \\
 $1$ & $[-0.75,-0.56,-0.33, -0.24,  -0.17,0.53, 0.57, 0.99]$ & $0.8$  \\
\end{tabular}
\caption{Three different parametrizations of the RBF networks and the corresponding system parameters for describing the lattice-regularized free scalar field theory in 1+1 dimensions.}
\label{tab:4}
\end{table}
Using the fitted parameters $a_k$, $c_k$, and $A$, the momentum space correlator can be calculated as:
\begin{equation}
\label{eq:O2}
\langle 0| \widetilde{\Phi}_{a,b} \widetilde{\Phi}_{c,d} | 0 \rangle = \frac{\prod \limits_{i,j}\int d\widetilde{\Phi}_{i,j} \; \widetilde{\Phi}_{a,b} \widetilde{\Phi}_{c,d} \; G(\widetilde{\Phi}_{i,j})  }{\prod \limits_{i,j}  \int d\widetilde{\Phi}_{i,j} \; G(\widetilde{\Phi}_{i,j})  } ,
\end{equation}
with
\begin{equation}
\label{eq:O1}
G(\widetilde{\Phi}_{i,j}) = \sum_{k=1}^K  a_k e^{   -\frac{1}{2} \Big[ \lambda_{ij}-2A\Big] \widetilde{\Phi}_{i,j}^2 + 2Ac_k \widetilde{\Phi}_{i,j} - A c_k^2 } ,
\end{equation}
where $ \widetilde{\Phi}_{i,j}$ is the value of the fields at the discretized momentum modes $(i,j)$, and $\lambda_{ij}$ are the eigenvalues of the discretized Laplace operator coming from the kinetic terms. The 2-point correlator in 1+1 dimensions is defined as the vacuum expectation value of the product of the fields at momentum modes $(a,b)$ and $(c,d)$ corresponding to momenta $p_{a,b}$ and $p_{c,d}$, and due to its quadratic Gaussian form, it can be integrated out in a closed form.

\begin{figure}[htbp]
    \centering
    \includegraphics[width=.4\textwidth]{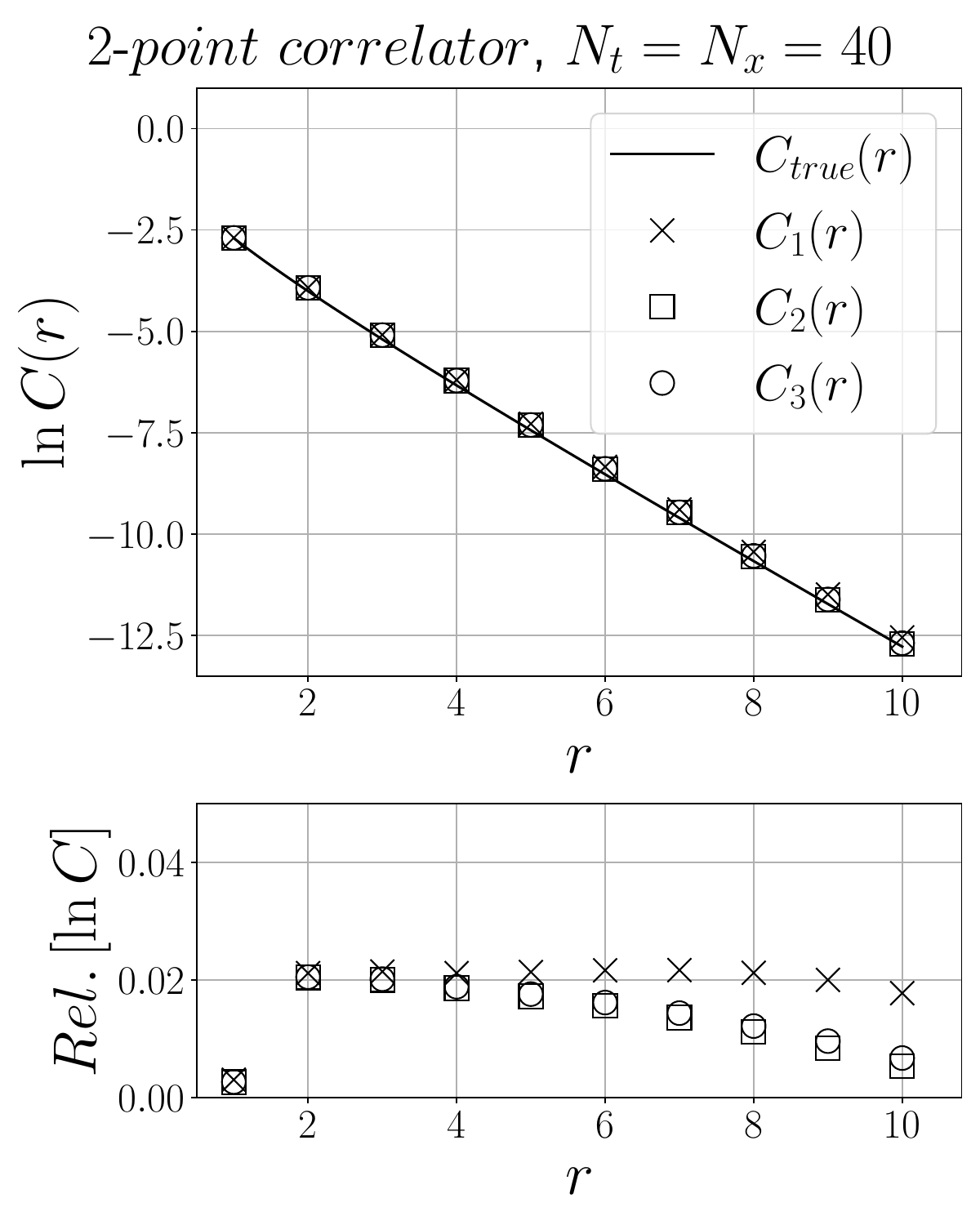}  
    \caption{
The logarithm of the true and RBF-approximated 2-point correlators in position space for the three different RBF network parametrizations. The lower panel shows the relative errors of $\ln C(r)$ for all the three cases.
}
    \label{fig:SC2}
\end{figure}

In addition, the scaling introduces an $S_c^2$ factor outside of the correlator due to the two field insertions, while $\lambda_{ij} \rightarrow S_c^2 \lambda_{ij}$ changes due to the scaled kinetic part. The momentum space correlator thus can be given as:
\begin{equation}
\label{eq:F4}
\widetilde{C}(p_t,p_x;q_t,q_x) = S_c^2 \langle 0 | \widetilde{\Phi}_{p_t,p_x} \widetilde{\Phi}_{q_t,q_x}| 0 \rangle_{\lambda \rightarrow S_c^2 \lambda},
\end{equation}
where $p_t, p_x, q_t, q_x$ are the corresponding momenta, where the correlator is evaluated. To get the coordinate space correlator, the inverse Fourier transform has to be applied to the momentum space correlator as:
\begin{eqnarray}
\label{eq:F5}
&&C(n_t,n_x;m_t,m_x) =\frac{1}{(N_t a) (N_x a)} \times \\
&&\sum_{p_t,p_x,q_t,q_x} e^{  2\pi i \left( \frac{p_t n_t+q_t m_t}{N_t}  + \frac{p_x n_x + q_x m_x}{N_x}  \right) } \widetilde{C}(p_t,p_x;q_t,q_x)  , \nonumber
\end{eqnarray}
where due to translational symmetry it is sufficient to fix the initial coordinates $m_t$ and $m_x$ to $0$. The exact solution for the coordinate space correlator in 1+1 dimensions can be given by the modified Bessel function of the second kind as \cite{51}:
\begin{equation}
\label{eq:F6}
C_{true}(r) = \frac{1}{2 \pi} K_1(mr),
\end{equation}
where the distance on the lattice is defined as $r=\sqrt{(n_xa-m_xa)^2 + (n_ta-m_ta)^2}$, while $K_1(mr)$ is the modified Bessel function of the second kind. In Fig.~\ref{fig:SC2} the true correlator is compared to the model results after inverse Fourier transforming the momentum space expressions, showing a very good match in all cases.
From the results shown here, we have seen that the true correlator can be reproduced with very good accuracy using the RBF approximated momentum space correlators. The actual parameters of the RBF kernels can be varied according to the function that has to be approximated, and the most important condition that has to be satisfied is that the centers have to be distributed approximately evenly around zero. By scaling the fields, the RBF parameters can be fine-tuned so that a good parametrization can be used.

In the next section, the numerical and time complexity of the RBF approximation will be addressed through examples of the free theory.

\section{ Numerical complexity of the RBF expansion }
\label{sec:22}
In this section, we will address the numerical and time complexity of the RBF method through the calculation of the coordinate space propagator of the free theory. All the calculations have been done on an LG Gram notebook with an Intel Core i5 processor, and 16 GB memory, without any direct GPU support or parallelisation.

In general, the time that it takes to calculate an observable with the model can be separated into two parts: (1) fitting the RBF network and (2) determining the observables from the momentum space expression shown in Eq.~\ref{eq:42}, thus, $T_{Tot}$ can be expressed as follows:
\begin{equation}
T_{Tot} = T_{RBF} + T_{O},
\end{equation}
where $T_{RBF}$ is the time of training, while $T_{O}$ includes everything that is needed to determine the specific observables.

In the case of real scalar fields, the time of training of the network ($T_{RBF}$) means the determination of the centers, width, and the $a_k$ weight parameters, and is essentially a one-dimensional problem, where the approximable function is given by $F[\phi]$ and the optimization can be written as:
\begin{equation}
\arg \min_{\!\!\!\!\!\!\!\!\!\!\!\! \{ a_k,c_k,A \}} \left\{ \frac{1}{N_T}\sum_{i=1}^{N_T} \left( F[\phi_i] - \sum_{k=1}^K a_k e^{-A(\phi_i-c_k)^2}\right)^2 \right\},
\end{equation}
with the main constraint that the centers need to be $\sim$symmetrically distributed around $\phi=0$. In practice, there are two main ways to train RBF networks: (1) fixing the number of centers and training all the parameters by, e.g., gradient descent or other efficient numerical techniques, (2) fixing the centers and widths, then solving the resulting linear least squares problem to the $a_k$ weight parameters. In this paper, we will apply the latter method, with some additional constraints included in the training procedure given by the following steps:
\\ \\
(1) Scale $F[\phi]$ into a predefined range using the $S_c>0$ scale parameter so that the dominant contribution of $F[\phi]$ will be given between $-L < \phi < L$.
\\ \\
(2) Place down $K$ kernels equidistantly between $c_k \in [-L,L]$ with a resolution of $\Delta c_k = \frac{2L}{K-1}$.
\\ \\
(3) Set up a grid of $A$ kernel widths and fit the $a_k$ RBF weights for each network with the given $c_k$ centers, and $A$ parameters. Train the network on a slightly larger interval, e.g., $\phi \in [-1.2\cdot L,1.2 \cdot L]$ using a randomly generated $N$ number of points, sampled in that interval.
\\ \\
(4) Do step (3) for increasing number of $K$ until some stopping condition is met, e.g., the mean squared error saturates.
\\ \\
Training on a slightly larger interval in step (3) makes sure that the network tries to generalize for the decay outside $[-L,L]$ as well. In practice, for a moderate number of centers, e.g., $K=10-100$, 1 training for one set of parameters can be done very fast in $\sim$ milliseconds. The complexity of this specific training procedure scales with the number of training samples $N_T$, the number of kernels $K$, and the number of iterations we do for different $A$'s and $K$'s, i.e., on the grid $(A_i,K_i)$ where we sweep through to find the optimal parameter set. Even with this 'brute-force' method, the training should not take more than a few seconds, however, the actual value depends on the parameter grid and the number of samples that are used for training.

\begin{figure}[htbp]
    \centering
    \includegraphics[width=.45\textwidth]{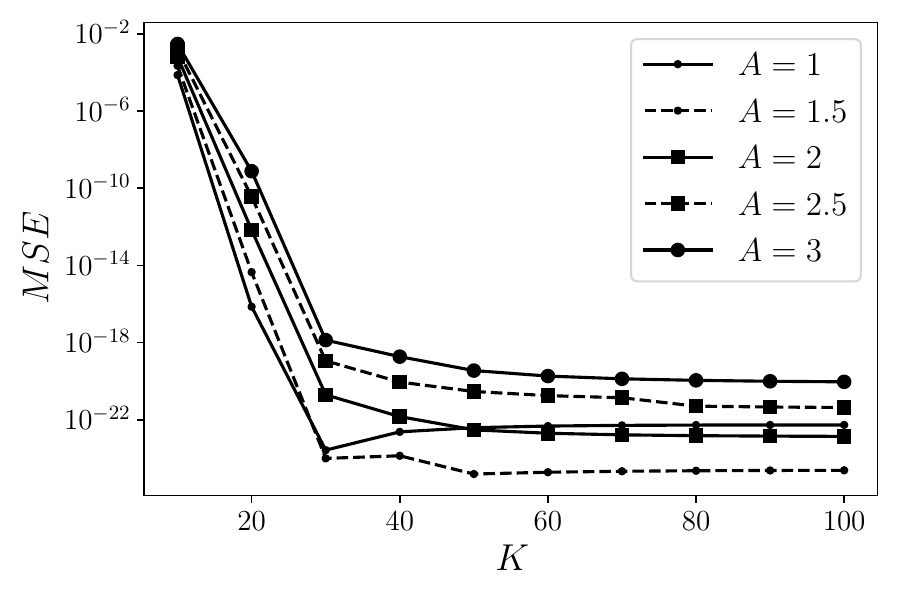}  
    \caption{
Mean squared errors of the training procedure for the free scalar field theory with $m=0.1$. In each case the scaling is set to $S_c=10$, and $N_T=1000$ samples are used between $\phi \in [-6,6]$ for training. The centers of each network are equidistantly spaced between $\phi \in [-5,5]$ with $\Delta c_k = 10/(K-1)$, while the width parameters are set to $A=[1,1.5,2,2.5,3]$.
}
    \label{fig:Ex2}
\end{figure}

\begin{figure}[htbp]
    \centering
    \includegraphics[width=.45\textwidth]{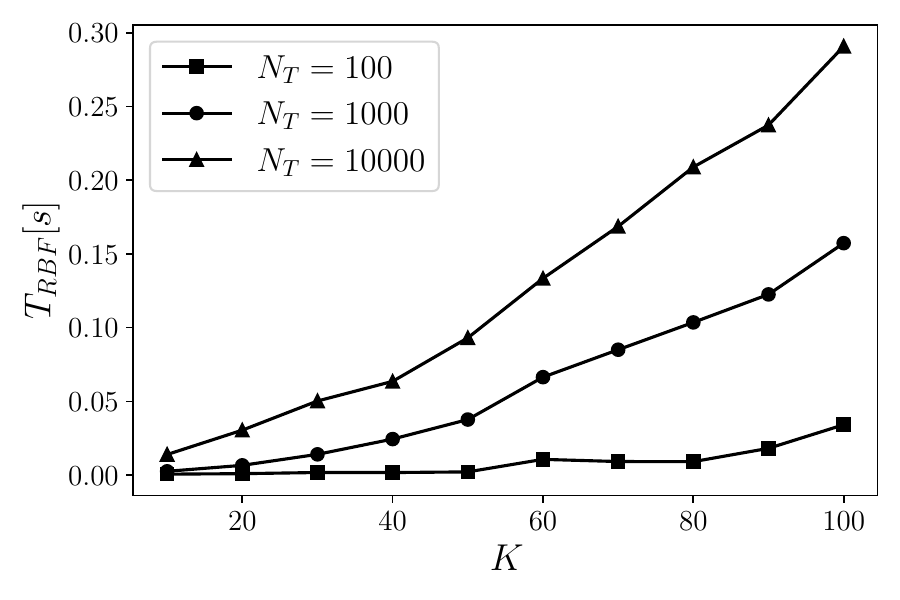}  
    \caption{
The time that it takes to train the RBF network ($T_{RBF}$) with different numbers of centers ($K$) and different numbers of training samples ($N_T$).
}
    \label{fig:Ex3}
\end{figure}

In Fig.~\ref{fig:Ex2}, we show the mean squared errors for a concrete example of free scalar fields with $m=0.1$ mass parameter. The scaling has been set to $S_c=10$, while the centers have been distributed between $\phi \in [-5,5]$ (thus setting $L=5$). Due to the reasoning of the previous section, we will try to dismiss a perfect Gaussian fit during training and set the grid of the width parameters as $A=[1,1.5,2,2.5,3]$, which will make sure that the original $F$ function is approximated by a mixture of narrower Gaussian kernel functions. The training has been done for each $A_i$ value with $K=[10,20,30,40,50,60,70,80,90,100]$ numbers of kernels and $N_T=1000$ numbers of samples. The results on Fig.~\ref{fig:Ex2} show that after $K=50$ the MSE does not change significantly, and the optimum is reached at $A=1.5$, and $K=50$ parameters.
The time of training ($T_{RBF}$) using the same (A,K) grid is shown in Fig.~\ref{fig:Ex3} for different $N_T$ and $K$ values. It can be seen that the time of training increases with increasing number of kernels, and increasing number of training samples, which is the expected behavior. The training for this system stays under a minute even for $N_T=10000$, and $K=100$. 

\begin{figure}[htbp]
    \centering
    \includegraphics[width=.45\textwidth]{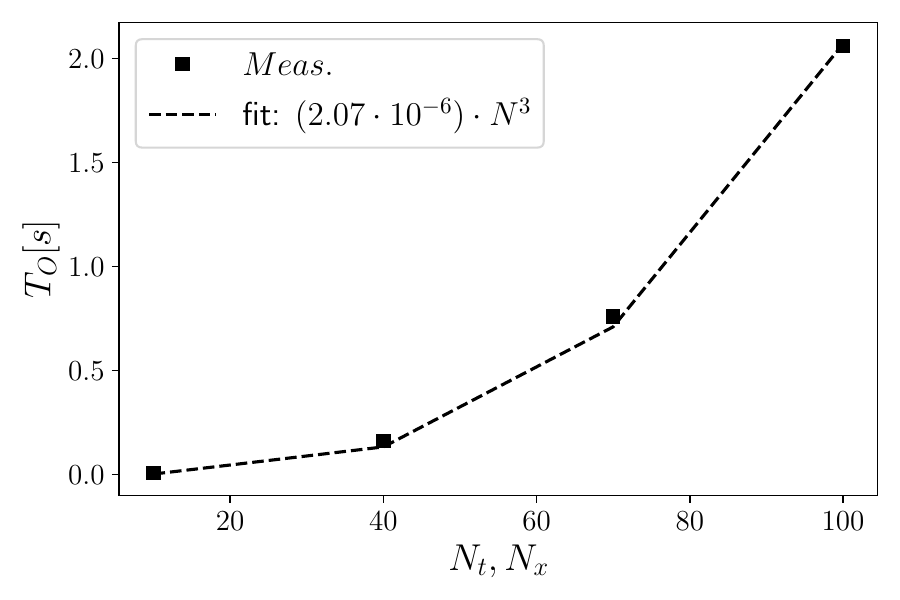}  
    \caption{
The time that it takes to calculate the coordinate space propagators on an $N_t \times N_x$ lattice after training the RBF network. In each case the propagator is calculated at $r=1,2,3,...N_t(N_x)$ coordinates (a total of $N_r=N_t=N_x$ points). The dashed line shows an $\mathcal{O}(N^3)$ fit to the measured values.
}
    \label{fig:Ex4}
\end{figure}

\begin{figure}[htbp]
    \centering
    \includegraphics[width=.45\textwidth]{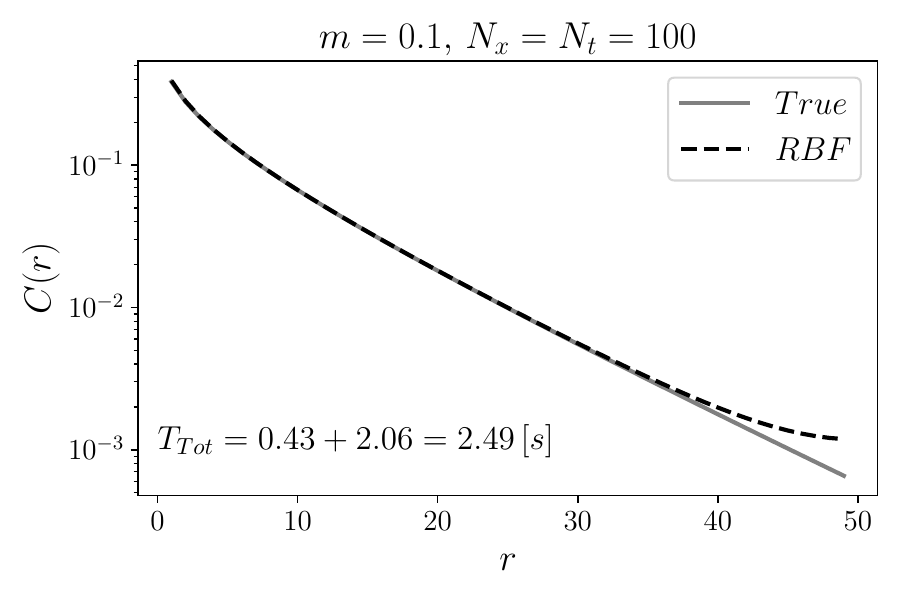}  
    \caption{
The true and RBF-approximated 2-point correlators in position space using the optimal RBF network with parameters $K=50$ and $A=1.5$ on a $100 \times 100$ lattice. The total time of obtaining the propagators with the RBF expansion is $T_{Tot}\approx 2.49$ [s].
}
    \label{fig:Ex5}
\end{figure}

Apart from the time that takes to train the network ($T_{RBF}$), we also need to include the time that takes to calculate the coordinate space propagator ($T_O$), starting from the momentum space expression of the path integral given by Eq.~\ref{eq:42}. This means the calculation of the momentum space propagator, then inverse Fourier transform it into coordinate space. The complexity of this depends on $N_T \times N_x$ size of the lattice, and in general, can be given by $\mathcal{O}(N_tN_xN_r)$, where $N_r$ is the number of coordinate space points, where we do the inverse Fourier transform. Let us set $N_r=N_x=N_t=10,40,70,100$ and calculate the coordinate space propagator for the same system with $m=0.1$. The results for $T_O$ can be followed in Fig.~\ref{fig:Ex4}, where a simple $N^3$ fit is also shown. From these results it can be seen that for large lattices, e.g., $100 \times 100$, the total time $T_{Tot}$ is dominated by the actual calculation of the observables, and not the training procedure itself.
In Fig.~\ref{fig:Ex5}, we also show the result for the coordinate space propagator together with the analytical result for a $100 \times 100$ lattice between $r \in [1,50]$. The total time that it takes to calculate the propagator (with $N_T=1000$ number of samples) is $T_{Tot}\approx 0.43+2.06=2.49$ [s], where the first value represents $T_{RBF}$, while the second represents $T_O$. A very good match is achieved until $r\approx 40$, where the boundary effects start to show themselves due to the periodic boundary conditions.

For comparison, to calculate the same propagator using, e.g., the Metropolis algorithm, it would need at least a few hundred thousand uncorrelated field configurations, which, depending on the autocorrelation length, would mean at least a few million samples. On the same notebook where the RBF calculations have been done, generating $10$ million samples using the Metropolis algorithm takes around $20$ minutes. After the generation of uncorrelated samples, the correlator has to be calculated, which also takes some time. 

The training procedure has been shown for the case of real scalar fields. In the case of complex scalar fields, $\phi_{i,j}$ can be separated into real and imaginary parts, thus, the input of the network needs another dimension, and the training procedure becomes a two-dimensional fit of the parameters on a two-dimensional grid of centers (see e.g., \cite{BG_FD}).

After the description of the time complexity, let us finish this section with two more additional numerical issues. In the previous example, we have set the mass parameter to $m=0.1$, and before that we have used $m=1$ to show the working principles of the model. In general, smaller masses mean larger correlation times, which could cause severe numerical problems in Monte Carlo techniques, e.g., critical slowing down near phase transitions. In the RBF expansion this should not pose additional problems (at least directly), because by setting the mass to very small values, it would only mean that the approximable function $F(\phi)$ changes shape depending on the Lagrangian density. In the case that the function scales to a larger effective region of $\phi$, by applying a corresponding scaling $S_c$, it can be rescaled into a more convenient range. This was shown in the previous example with $m=0.1$ and $S_c=10$. In this case the method still remains the same, i.e., approximate the $F(\phi)$ function with an appropriate RBF network, which, depending on the system, might need more or fewer kernels or larger or smaller width parameters, but ultimately it does not require more sophisticated training methods.

\begin{figure}[htbp]
    \centering
    \includegraphics[width=.47\textwidth]{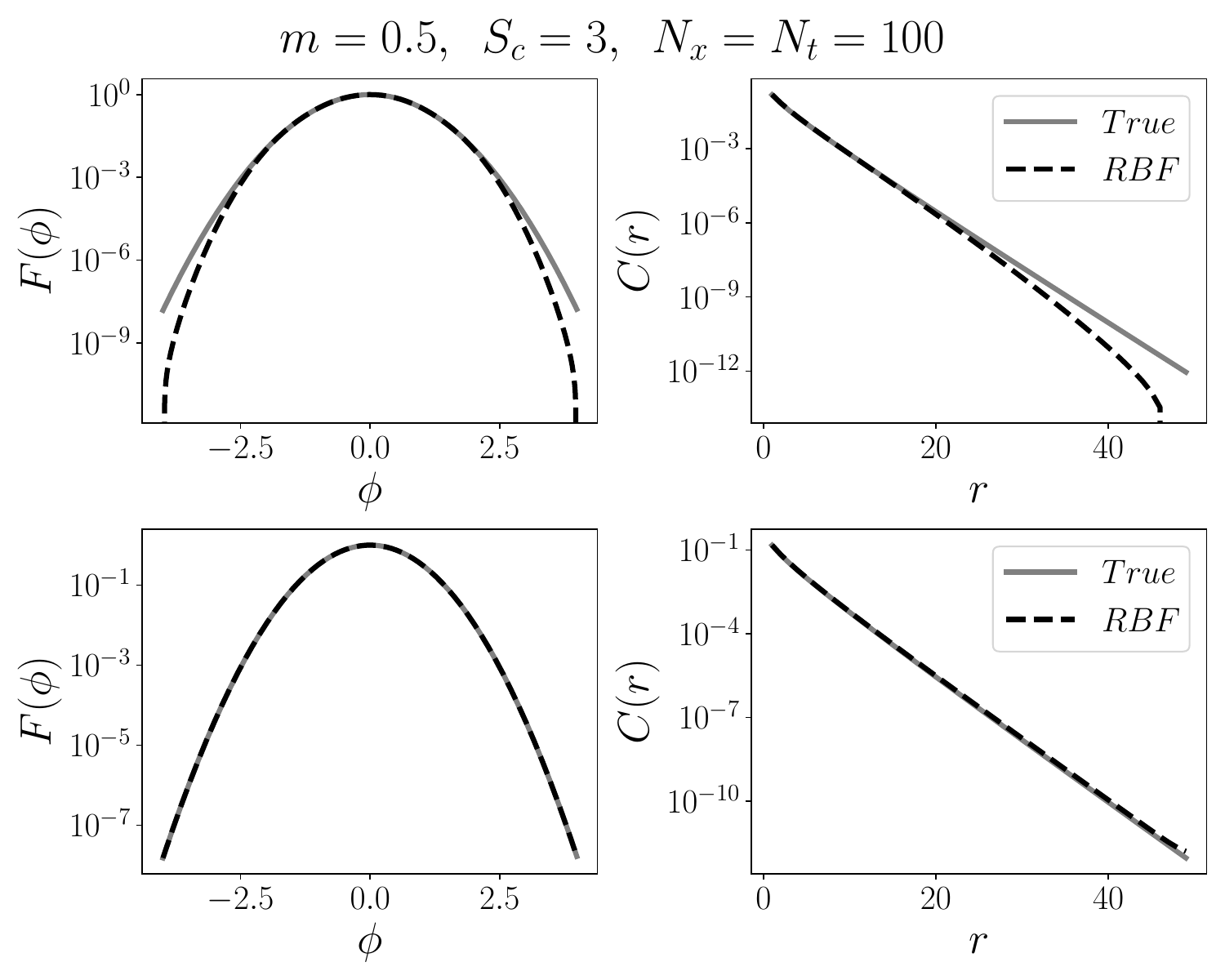}  
    \caption{
The RBF approximation of the $F(\phi)$ function and the correlators in the case of $m=0.5$, using two different parametrizations of the RBF networks. In both cases the scaling is set to $S_c=3$, and $A=2$ width parameter is used. For training, $N_T=1000$ samples were generated in the interval of $\phi \in [-4,4]$, while the centers are given by (1) $c_k = [-1.2, -0.9, -0.6, -0.3, 0.3, 0.6, 0.9, 1.2]$ (upper), and (2) $c_k \in [-3,3]$, with $K=20$ (lower).
}
    \label{fig:m05N100}
\end{figure}

The second numerical issue is regarding the asymptotic behavior of the system, i.e., what happens when we include larger $\phi$ values in the simulations. At its core, the RBF approximation cannot fully extrapolate to the regions where there were no training samples, therefore, a perfect fit for all $\phi$ values is not possible. This is, however, generally not needed, because even in Monte Carlo techniques the sampled $\phi$ values should have finite limits, where the dominant region of the system is sampled. This is the same in the RBF expansion, where we assume that the outside regions have negligible contributions to the observables in question. Determining the dominant region is a modelling step that has to be carefully examined. In the case of coordinate space correlators (as in the previous examples), larger masses and larger lattices will necessarily have very small numerical values at larger $r$ distances, therefore, a more precise approximation of $F(\phi)$ is needed. To show this, we have made two calculations on a $100\times100$ lattice for the free system with $m=0.5$. The scaling parameter has been set to $S_c=3$, and the training samples have been generated in the interval between $\phi \in [-4,4]$. Two RBF networks have been trained with the same widths, $A=2$, and centers given by:
\\ \\
(1) $c_k = [-1.2, -0.9, -0.6, -0.3,  0.3,  0.6,  0.9,  1.2]$
\\ \\ 
(2) $c_k \in [-3,3]$, with $K=20$, thus $\Delta c_k = \frac{6}{19}$ (centers placed equidistantly).
\\ \\
As before, both centers omit the trivial $c_k=0$, and $A=\frac{S_c^2 a^2 m^2}{2} = 1.125$ kernel, thus, the RBF network will approximate $F(\phi)$ with a mixture of Gaussians. Also note that centers of both networks are smaller than the range of $\phi$ where the training takes place. The results for the approximation of $F(\phi)$ and the $C(r)$ correlator can be followed in Fig.~\ref{fig:m05N100}.

From the result, some important consequences can be seen regarding the RBF approximation of $F(\phi)$ at very small (nondominant) values. The upper panel shows the case of network (1), in which case, while a dominant part of $F(\phi)$ is well approximated, the network underestimates the values outside $|\phi|\gtrsim2.5$, where $F(\phi)\lesssim10^{-5}$. This will cause an underestimation of the correlator after $r\gtrsim20$, where $C(r) \lesssim 10^{-6}$. In the case of an overestimation of $F(\phi)$, the correlator also tends to be overestimated at around the same values. In contrast, in the case of network (2) with centers placed equidistantly between $c_k \in [-3,3]$, both $F(\phi)$ and thus the correlator $C(r)$ are well approximated at all $r$ coordinates in the range that is allowed by the size of the lattice, and the periodic boundary conditions.

Due to the fact that the RBF approximation cannot follow the full asymptotics of any $F(\phi)$ function, it is an important modeling step to approximate the effective range of the input variable $\phi$, where an observable can be calculated with good accuracy. In practice this can be done by systematically calculating the observables at larger and larger intervals, which could mean different RBF networks with different parameters, e.g., more centers, larger widths, etc., until convergence is reached in the specific observable.

In the next section the interacting $\phi^4$ theory will be addressed by determining the phase transition line that separates the broken and unbroken phases.
\section{$\phi^4$ theory in 1+1 dimensions}
\label{sec:3}
The interacting scalar $\phi^4$ theory is one of the simplest quantum field theories, where the nonperturbative effects, arising from the quartic self-interactions, can be analyzed in a relatively simple manner. Despite its simple form, the nonlinear quartic self-interaction makes it possible to study phase transitions, critical phenomena, and spontaneous symmetry breaking and can serve as a prototype model to understand more complex systems, e.g., in condensed matter physics, cosmology, or the standard model \cite{53,54}. In this section the RBF model will be applied to the 1+1 dimensional real scalar $\phi^4$ theory and show that it is capable of describing the discretized system with an accuracy compared to previous lattice results.

The calculations that will be shown here do not aim to give a full, comprehensive description of the system (e.g., the continuum limit is not considered, or no finite size scaling is examined), but they are aimed at showing that the method is able to calculate the necessary observables to describe renormalized parameters and phase transitions. Due to the well-defined form of the RBF approximation, these extensions, e.g., taking the continuum limit, should not pose any problems and could be done by using the same techniques as it is done, for example, in lattice methods \cite{55}. One of the main advantages of the RBF method in comparison to lattice Monte Carlo methods is that the calculations can be done very fast, e.g., obtaining the phase transition line for multiple parameter combinations only takes a few seconds on a standard notebook, which would otherwise take at least many hours or even days with lattice Monte Carlo methods. The other advantages and future prospects of the RBF model will be summarized later in the concluding section.

In the D-dimensional continuum theory, the (Euclidean) action integral of the interacting $\phi^4$ model can be written in the following form:
\begin{equation}
\label{eq:P1}
S=\int d^Dx \; \left( \frac{1}{2} \partial_\mu \phi(x) \, \partial^\mu \phi(x) + \frac{m_0^2}{2} \phi(x)^2 + \frac{\lambda_0}{4!}\phi(x)^4 \right),
\end{equation}
where $m_0$ represents the bare mass, while $\lambda_0$ is the bare coupling that controls the strength of the quartic self-interaction. The discretized action defined from the continuum theory depends on the dimensionless $(am_0)^2$ and $(a^2\lambda_0)$ parameters, where both $m_0^2$ and $\lambda_0$ have dimensions of mass squared.

In this work, we only consider the lattice-regularized theory with finite $a$, and will not take the continuum limit that is otherwise needed to compare the calculations to experimental results. This would require a careful extrapolation and the calculation of the observables using different (decreasing) lattice spacings by keeping some corresponding parameter combinations fixed. One of the main difficulties using lattice Monte Carlo methods is the very long time that it takes to do the simulations at small lattice spacings, therefore, usually only a few different combinations are examined and then extrapolated to $a=0$. In some future works, we aim to address this issue as well, but for now we will stay with the lattice regularized theory with finite $a$ to be able to compare our results to the Monte Carlo calculations.

In this discretized setting, the function that has to be approximated by the RBF network can be written as follows:
\begin{equation}
\label{eq:P3}
F(\phi_{i,j}) = e^{ - S_c^2\frac{(a m_0)^2}{2}\phi_{i,j}^2   - S_c^4\frac{(a^2 \lambda_0)}{4!}\phi_{i,j}^4 } ,
\end{equation}
where the $S_c$ field scaling is also included in the description.
In the calculations that will be shown here, we have set the lattice spacing to $a=1$ and the lattice size to $N_t=N_x=100$. 

The Lagrangian of the model, with positive mass squared $m_0^2 \geq 0$ and positive coupling $\lambda_0 > 0$, suggests that the system is in the unbroken phase, where the positive coupling makes sure that the potential is bounded from below, which is necessary for stability. The system in this phase possesses a global $\mathbb{Z}_2$ symmetry, and the corresponding potential has a stable global minimum at zero vacuum expectation value. 
In the case of a positive coupling $\lambda>0$ and some corresponding negative mass-squared, the system undergoes spontaneous symmetry breaking, in which case the vacuum state no longer respects the original $\mathbb{Z}_2$ symmetry of the Lagrangian. 
Instead of a single global minimum at the origin, the potential develops a set of new minima away from zero, which leads to nonzero vacuum expectation values for the field, that is a general sign of the spontaneously broken symmetry.

In this state the system settles into one of the minima, thus selecting a specific vacuum state that breaks the symmetry of the original theory. Due to field fluctuations and finite mass shifts, the phase transition does not happen strictly when $m_0^2$ becomes negative, and without precise numerical methods that could tackle the nonperturbative regions (e.g., in the strong coupling limit), it is not straightforward to see what would be the critical points where phase transition occurs. One usual way on how the transition point is looked for (e.g. in lattice Monte Carlo methods) is to examine the behavior of the renormalized masses with the change of the coupling \cite{P2}. A general sign in this case would be the fast decrease of the mass at the critical coupling. Another signal that is usually searched for is the point where the vacuum expectation value of the fields becomes finite, which signals that the system settles into one of the new vacuum states. This, however, is not always trivial to do without any extra steps, because in the $\phi^4$ theory the new vacuum state is also symmetric, therefore, taking the expectation value averages out the fluctuations to zero \cite{P3}. There are several methods that exist on how to handle this issue on the lattice, e.g., by introducing an explicit breaking of the symmetry, thus forcing the system to jump into one of the minima, or by a careful construction of how the vacuum expectation value is taken \cite{61}. Here, we will use a different technique based on the effective potential approach that is more suitable for the RBF model.

To define the effective potential, first we have to define the effective action, that is, the quantum-corrected version of the classical action and serves as a generating functional for the one-particle irreducible correlation functions of the theory \cite{P4}. First, let us assume an $x$-dependent source term $J(x)$ and define the classical field $\phi_c$ that is probed by the background source term as follows:
\begin{equation}
\label{eq:J1}
\phi_c \equiv \frac{\delta W[J]}{\delta J(x)} ,
\end{equation}
where $W[J]=\ln Z[J]$ is the generating functional of the connected correlators. The effective action that governs the quantum-corrected dynamics of the classical field $\phi_c$ can be defined using the Legendre transform of the $W[J]$ functional as:
\begin{equation}
\label{eq:J2}
\Gamma[\phi_c] = W[J] - \int d^Dx \; J(x) \phi_c(x),
\end{equation}
where $\Gamma[\phi_c]$ is understood as the functional of the classical field configurations. The effective potential \cite{P5} captures the quantum corrections to the classical potential and can be given using the effective action by assuming a constant background field $J$. This static part of the effective action does not govern any dynamics and is useful to extract the vacuum structure and to determine the true vacuum of the theory by minimizing $V_{eff}(\phi)$. By probing the system with a set of different $J$ background fields, the vacuum expectation value of the field will be a function of $J$. By inverting this relation, the effective potential can be expressed as follows:
\begin{equation}
\label{eq:J3}
V_{eff}(\phi) \propto \int_0^{\phi} d \langle \phi \rangle \; J (\langle \phi \rangle) ,
\end{equation}
where the background field $J$ is understood as a function of the vacuum expectation value of the fields, while due to the constant background field, $\langle \phi \rangle$ can be calculated by taking the derivative of the generating functional with respect to the field shift as:
\begin{equation}
\langle \phi \rangle_{J_0} = \frac{\delta \ln Z[J]}{\delta J} \Big|_{J=J_0},
\end{equation}
where $Z$ is given by the RBF expansion of the partition function defined in Eq.~\ref{eq:41}. 

The method we will follow starts by calculating the $J$ dependence of the vacuum expectation of the averaged fields $\langle \phi \rangle_J$ for a range of $J$ background fields. To do this, let us define the shifted $F(\phi_{i,j},J)$ function as:
\begin{equation}
\label{eq:J4}
F(\phi_{i,j},J) = e^{ - S_c^2\frac{(a m_0)^2}{2}\phi_{i,j}^2   - S_c^4\frac{(a^2 \lambda_0)}{4!}\phi_{i,j}^4 + J S_c \phi_{i,j} } ,
\end{equation}
where $J$ is a constant that represents the background field, and the corresponding extra $J\phi_{i,j}$ term is also scaled by $S_c$ due to the field scaling. 

In this approach, by calculating $\langle \phi \rangle_J$ for a set of $J$ fields, we can determine the $J(\langle \phi \rangle)$ from which the effective potential $V_{eff}$ could be determined. This method, however, is not that straightforward due to the properties of the Legendre transform, which requires convexity from the functions involved \cite{P6,P7}. If the potential is a nonconvex function, then the Legendre transform will correspond to its convex hull, which will be the case in the broken phase, where the potential includes many local minima and is a nonconvex function of $\phi$. The arising problems due to the nonconvexity in the broken phase are addressed in \cite{P8}, where it is indicated that the effective potential that can be obtained by lattice Monte Carlo simulations correspond to the Maxwell-construction between its saddle points. 
In practice this will correspond to straight lines that connect the local minima of the potentials, thus making the effective potential a convex function of $\langle \phi \rangle$. The same behavior is also described in \cite{P9} for a more general case, to study of first- and second-order phase transitions.

The results shown here will have the same behavior for the lattice-regularized $\phi^4$ theory, therefore, we need a prescription on how to handle this issue. One way would be to identify the end points of the linear parts that correspond to the Maxwell construction and extract the corresponding vacuum expectation values. This is, however, not the most precise way to do so, because of the uncertainties of the results that could distort the obtained points near $J=0$, which would be necessary to know with very good accuracy. A better and more robust method is to make a simple Ansatz for the effective potential as follows \cite{P10}:
\begin{equation}
\label{eq:J5}
V_{eff}(\langle \phi \rangle) = \hat{A}(m_R,Z_R) \langle \phi \rangle^2 + \hat{B}(\lambda_R,Z_R) \langle \phi \rangle^4 ,
\end{equation}
where the $A(m_R,Z_R)$ parameter depends on the renormalized mass $m_R$ and the field renormalization constant $Z_R$, while the $B(\lambda_R,Z_R)$ parameter also depends on the renormalized coupling $\lambda_R$. 
In a more general case, higher order and logarithmic terms could also be included in the effective potential, however, in this case, this simple functional form is enough, especially in the large $J$ region where we are mostly interested in. Using this general form for the effective potential makes it possible to make a robust fit to the whole range of the obtained $J(\langle \phi \rangle)$ points through the following parametrization:
\begin{equation}
\label{eq:J6}
J(\langle \phi \rangle) = a\langle \phi \rangle^3+ b \langle \phi \rangle \; \; \rightarrow \; \; 
\begin{cases}
	\hat{A}(m_R,Z_R) = b/2 \\
	\hat{B}(\lambda_R,Z_R)=a/4
\end{cases}
\end{equation}
in which case the minima of the potential can be expressed using the fitted $(a,b)$ parameters as:
\begin{equation}
\label{eq:J7}
\langle \phi \rangle_{min} = 
\begin{cases}
	0,  & b \geq 0 \\
	\pm \sqrt{\frac{|b|}{a}}, & b<0
\end{cases}
\end{equation}
where in the nonbroken phase we will have $b>0$, while in the broken phase $b<0$. Using the Ansatz in Eq.~\ref{eq:J5} will correspond to the Maxwell construction of the potential in the case of the $\phi^4$ theory due to its symmetry, therefore, a good choice to study the phase transition points of the theory. The full form of $V_{eff}(m_R,Z_R,\lambda_R)$ would also make it possible to extract the other renormalized parameters like $\lambda_R$ and $Z_R$, if $m_R$ is known from, e.g., studying the momentum space correlator at small $p^2$, or by fitting the exponential decay of the coordinate space correlators at large (Euclidean) times.

\begin{figure}
    \centering
    \includegraphics[width=.48\textwidth]{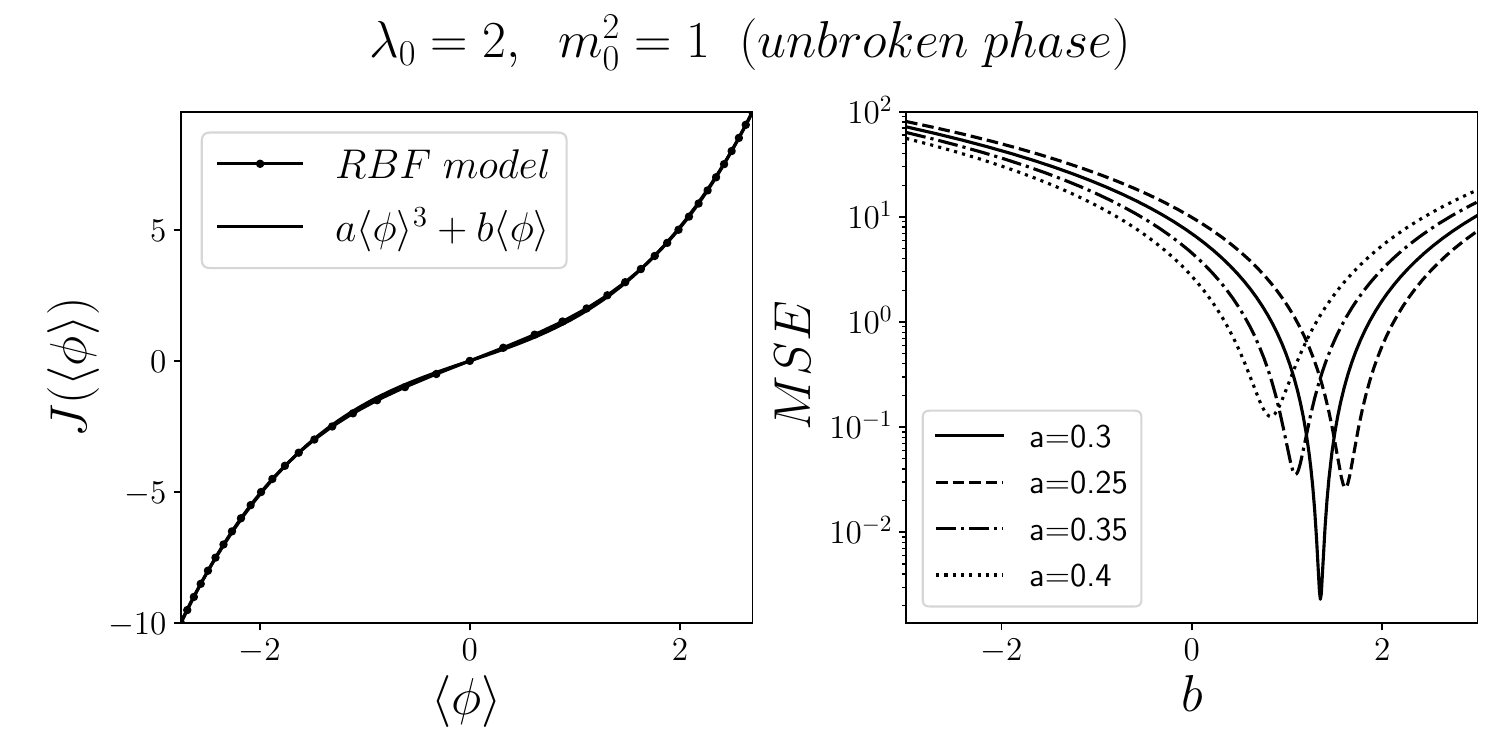}  
    \caption{The obtained $J(\langle \phi \rangle)$ functional form calculated by the RBF model for a set of $J$ background fields and the corresponding fit using Eq.~\ref{eq:J6} for the unbroken phase. The right hand side plot shows the mean squared error for different $(a,b)$ combinations. The minimum of the mean squared error corresponds to $b>0$, which signals that the system is in the unbroken phase. }
    \label{fig:PH1}
\end{figure}

\begin{figure}
    \centering
    \includegraphics[width=.48\textwidth]{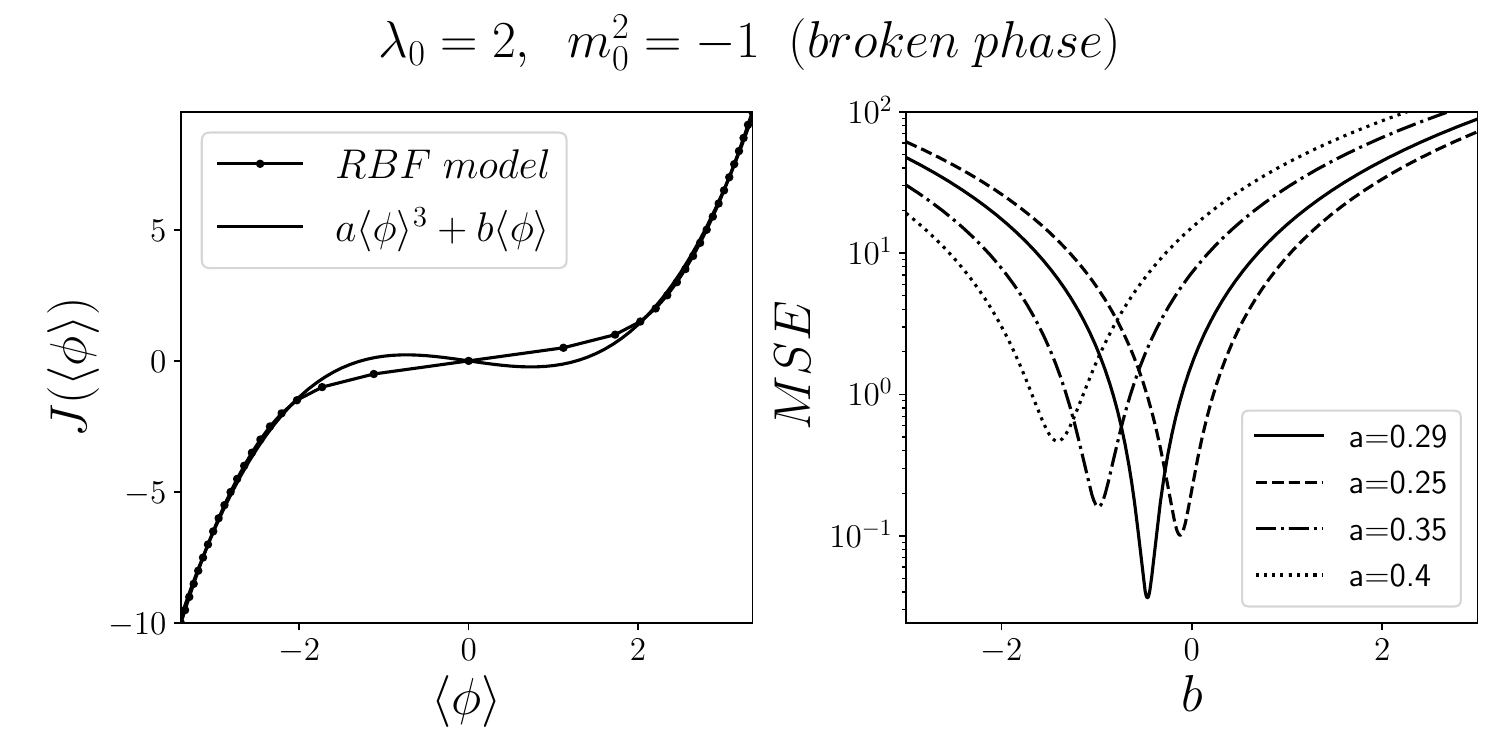}  
    \caption{The obtained $J(\langle \phi \rangle)$ functional form calculated by the RBF model for a set of $J$ background fields and the corresponding fit using Eq.~\ref{eq:J6} for the broken phase. The right hand side plot shows the mean squared error for different $(a,b)$ combinations. The minimum of the mean squared error corresponds to $b<0$, which signals that the system is in the broken phase.}
    \label{fig:PH2}
\end{figure}

In Fig.~\ref{fig:PH1} and in Fig.~\ref{fig:PH2}, two examples are shown for the unbroken and for the broken phase, respectively, where on the left side the obtained $J(\langle \phi \rangle)$ points (by using the RBF approximation) and the corresponding fit using Eq.~\ref{eq:J6} are shown. In both cases the right side plots show the mean squared errors (MSE) for different $(a,b)$ combinations.
From Fig.~\ref{fig:PH2}, which corresponds to the broken phase, the previously described convexity problem is clearly visible through observing the discrete points coming from the numerical results. In this case the part between $\langle \phi \rangle \approx [-1.2,1.2]$ indicates that the phase is broken and there would be two minima with finite vacuum expectation values. 

\begin{figure}
    \centering
    \includegraphics[width=0.5\textwidth]{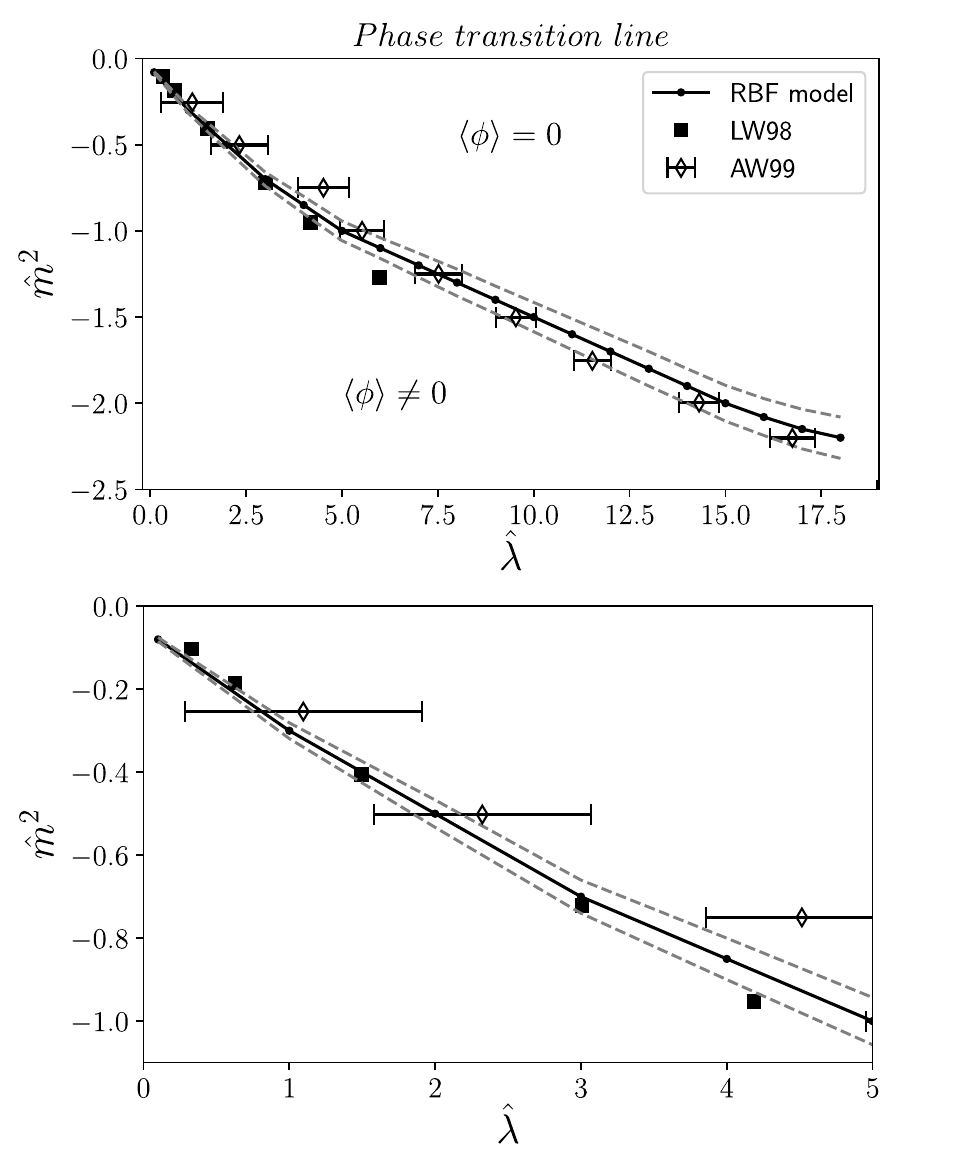}  
    \caption{Phase transition line of the $\phi^4$ theory in 1+1 dimensions that is obtained from the RBF expansion and compared to the results from lattice Monte Carlo simulations from LW98 \cite{P11} and AW99 \cite{P12}. The lower figure shows the same results as the upper panel but gives more focus to the range of small $\hat{m}$ values.
}
    \label{fig:PH3}
\end{figure}

Considering all the above, the task that is to find the transition point that corresponds to the phase transition of the $\phi^4$ theory, can be reformulated into finding the corresponding $m_0$ parameter for a fixed $\lambda_0$ coupling, where the best fit for the $J(\langle \phi \rangle)$ function (in the least square sense) has a transition from positive to negative $b$ values (unbroken $\rightarrow$ broken phase), thus, we are searching for the $b=0$ transition point.
In Fig.~\ref{fig:PH3}, the RBF results for the phase transition points are shown together with previous Monte Carlo simulations on the lattice, taken from [LW98] \cite{P11} and [AW99] \cite{P12}. The Monte Carlo data that are used for the comparison show good agreement with each other, and even though [LW98] does not have associated error bars, its data have been used to estimate the critical coupling in the continuum limit with a few percent accuracy.
The comparison between the RBF expansion and the two Monte Carlo results shows a very good match in a wide range of $(a^2\lambda_0)$ coupling strengths.
 
The uncertainty of the RBF results for each $(\hat{m}^2,\hat{\lambda})$ pair has been estimated by calculating the standard deviation of the fitted $b$ parameter from the covariance matrices of the least squares estimate, then propagate the error to obtain the standard deviation of the $m_0$ parameter at fixed $\hat{\lambda}$, at the transition point, where $b=0$. 

Considering our results, it seems that the effective potential approach is sufficient to explain the behavior of the phase transition line in the given coupling region. The next step in this direction would be to take the continuum limit and calculating the $f_{c}=\lambda_R/m_R^2$ critical coupling that characterizes the transition between the unbroken and broken phases. 

\section{Conclusions}
\label{sec:4}
In this paper a neural network based method is proposed to solve Euclidean path integrals for interacting systems in quantum field theories. The method is based on a radial basis function-type neural network expansion of the nonlinear interacting terms in the path integral, in which case the full discretized path integral can be expressed in a closed, analytically tractable form. The method makes it possible to calculate the generating functions, masses, vacuum expectation values, and other observables in a very fast and compact way that greatly exceeds the capabilities of other methods that aim to solve the nonperturbative regions of quantum field theories. The model has been tested for the lattice regularized interacting (real) scalar $\phi^4$ theory, where the phase transition from unbroken to broken phase have been studied. To determine the phase transition line, an effective potential approach is used, where the vacuum expectation values of the averaged fields have been determined for a set of finite background field shifts, then, by examining the effective potential, the transition points were determined. The results have been compared to previous lattice calculations, giving a very good agreement between them. 
Taking the thermodynamic limit is straightforward, as in the case of any lattice technique, by doing the calculations at larger volumes and then extrapolating to the infinite-size limit. On the other hand, taking the continuum limit is generally not a trivial task and requires a careful examination and scaling of the system at smaller and smaller resolutions. 

The generality of the RBF model makes it possible to estimate not just real but imaginary functional forms as well, in which case the path integral could be cast into a sum of generalized quadratic path integrals having real and imaginary parts. This representation could overcome the difficulties that arise e.g. in finite density calculations using Monte Carlo techniques. 
The main advantage of the method at this stage is its speed and capability to study the nonperturbative aspects of interacting field theories. With some straightforward extensions, there is a possibility that in the future it could also be able to describe non-Abelian gauge theories at finite densities, which is a necessary step in, e.g., describing the phase structure or vacuum properties of quantum chromodynamics and the strongly interacting matter.

\begin{acknowledgments}
This work was supported by the Korea National Research Foundation under Grant No. 2023R1A2C300302311 and 2023K2A9A1A0609492411, and the Hungarian OTKA fund K138277.
\end{acknowledgments}


\end{document}